\documentclass[conference]{IEEEtran}
\makeatletter
\def\ps@headings{%
\def\@oddhead{\mbox{}\scriptsize\rightmark \hfil \thepage}%
\def\@evenhead{\scriptsize\thepage \hfil \leftmark\mbox{}}%
\def\@oddfoot{}%
\def\@evenfoot{}}
\makeatother
\ifCLASSOPTIONcompsoc
  \usepackage[caption=false,font=normalsize,labelfont=sf,textfont=sf]{subfig}
\else
  \usepackage[caption=false,font=footnotesize]{subfig}
\fi

\usepackage{graphicx, amsmath, amssymb, subfig, setspace}
\usepackage{algorithm}
\usepackage{algorithmic} 
\usepackage{url}
\usepackage{multirow}

\newtheorem{proposition}{\bf Proposition}
\newtheorem{corollary}{\bf Corollary}
\newtheorem{lemma}{\bf Lemma}
\newtheorem{definition}{\bf Definition}

\def\squareforqed{\IEEEQED}
\def\qed{\ifmmode\squareforqed\else{\unskip\nobreak\hfil
\penalty50\hskip1em\null\nobreak\hfil\squareforqed
\parfillskip=0pt\finalhyphendemerits=0\endgraf}\fi}

\def\hollowsquare{\hbox{\rlap{$\sqcap$}$\sqcup$}}
\def\hollowqed{\ifmmode\hollowsquare\else{\unskip\nobreak\hfil
\penalty50\hskip1em\null\nobreak\hfil\hollowsquare
\parfillskip=0pt\finalhyphendemerits=0\endgraf}\fi}

\newcommand{\ud}{\mathrm{d}}
\newcommand{\E}{\mathbf{E}}

\newcommand{\s}{\mathbf{s}}

\begin{document}
	
\title{
  To Reserve or Not to Reserve: Optimal Online Multi-Instance Acquisition in IaaS Clouds
  }
\author{
\IEEEauthorblockN{Wei Wang, Baochun Li, and Ben Liang}
\IEEEauthorblockA{Department of Electrical and Computer Engineering\\
University of Toronto}
\vspace*{-0.3in}
}
	
\maketitle

\pagestyle{plain}
\thispagestyle{plain}

\begin{abstract}

Infrastructure-as-a-Service (IaaS) clouds offer diverse instance purchasing options. A user can either run instances on demand and pay only for what it uses, or it can prepay to reserve instances for a long period, during which a usage discount is entitled. An important problem facing a user is how these two instance options can be dynamically combined to serve time-varying demands at minimum cost. Existing strategies in the literature, however, require either exact knowledge or the distribution of demands in the long-term future, which significantly limits their use in practice.  Unlike existing works, we propose two practical {\em online algorithms}, one deterministic and another randomized, that dynamically combine the two instance options online without any knowledge of the future. We show that the proposed deterministic (resp., randomized) algorithm incurs no more than $2-\alpha$ (resp., $e/(e-1+\alpha)$) times the minimum cost obtained by an optimal {\em offline algorithm} that knows the exact future {\em a priori}, where $\alpha$ is the entitled discount after reservation. Our online algorithms achieve the best possible competitive ratios in both the deterministic and randomized cases, and can be easily extended to cases when short-term predictions are reliable. Simulations driven by a large volume of real-world traces show that significant cost savings can be achieved with prevalent IaaS prices.

\end{abstract}
	
\section{Introduction}
\label{sec:intro}

Enterprise spending on Infrastructure-as-a-Service (IaaS) cloud is on a rapid
growth path. According to \cite{gartner12}, the public cloud services market is
expected to expand from \$109 billion in 2012 to \$207 billion by 2016, during
which IaaS is the fastest-growing segment with a 41.7\% annual growing rate
\cite{cloudtimes12}. IaaS cost management therefore receives significant
attention and has become a primary concern for IT enterprises.

Maintaining optimal cost management is especially challenging, given the
complex pricing options offered in today's IaaS services market. IaaS cloud
vendors, such as Amazon EC2, ElasticHosts, GoGrid, etc., apply diverse instance
(i.e., virtual machine) pricing models at different commitment levels. At the
lowest level, cloud users launch {\em on-demand} instances and pay only for the
incurred instance-hours, without making any long-term usage commitments, e.g.,
\cite{ec2pricing, elastichosts, gogrid}.  At a higher level, there are {\em
reserved instances} wherein users prepay a one-time upfront fee and then
reserve an instance for months or years, during which the usage is either free,
e.g., \cite{elastichosts, gogrid}, or is priced under a significant discount,
e.g., \cite{ec2pricing}. Table~\ref{tbl:ec2pricing} gives a pricing example of
on-demand and reserved instances in Amazon EC2.

Acquiring instances at the cost-optimal commitment level plays a central role
for cost management. Simply operating the entire load with on-demand instances
can be highly inefficient. For example, in Amazon EC2, three years of
continuous on-demand service cost 3 times more than reserving instances for
the same period \cite{ec2pricing}. On the other hand, naively switching to a
long-term commitment incurs a huge amount of upfront payment (more than 1,000
times the on-demand rate in EC2 \cite{ec2pricing}), making reserved instances
extremely expensive for sporadic workload. In particular, with time-varying
loads, a user needs to answer two important questions: (1) when should I
reserve instances (timing), and (2) how many instances should I reserve
(quantity)?



\begin{table}[t]
   \centering
   \renewcommand{\arraystretch}{0.99}
   \footnotesize
   \caption{Pricing of on-demand and reserved instances (Light Utilization,
   Linux, US East) in Amazon EC2, as of Feb. 10, 2013.}
   \vspace{-2mm}
   \begin{tabular}{|c||c|c|c|}
     \hline
     {\bf Instance Type} & {\bf Pricing Option} & {\bf Upfront} & {\bf Hourly} \\
     \hline
     \multirow{2}{*}{Standard Small} & On-Demand & \$0 & \$0.08 \\ \cline{2-4}
     & 1-Year Reserved & \$69 & \$0.039 \\
     \hline
     \multirow{2}{*}{Standard Medium} & On-Demand & \$0 & \$0.16 \\ \cline{2-4}
     & 1-Year Reserved & \$138 & \$0.078 \\
     \hline
   \end{tabular}
   \label{tbl:ec2pricing}
   \vspace{-.1in}
\end{table}

Recently proposed instance reservation strategies, e.g., \cite{hong11,
bodenstein11, verm11}, heavily rely on long-term predictions of future demands,
with historic workloads as references. These approaches, however, suffer from
several significant limitations in practice. First, historic workloads might
not be available, especially for startup companies who have just switched to
IaaS services. In addition, not all workloads are amenable to prediction.
In fact, it is observed in real production applications that workload is highly
variable and statistically nonstationary \cite{stewart07, singh10}, and as a
result, history may reveal very little information about the future. Moreover, due
to the long span of a reservation period (e.g., 1 to 3 years in Amazon EC2),
workload predictions are usually required over a very long period of time, say,
years. It would be very challenging, if not impossible, to make sufficiently
accurate predictions over such a long term.  For all these reasons, instance
reservations are usually made conservatively in practice, based on empirical
experiences \cite{aws-case-study} or professional recommendations, e.g.,
\cite{cloudability, cloudyn, cloudexpress}.



In this paper, we are motivated by a practical yet fundamental question: Is it
possible to reserve instances in an {\em online} manner, with limited or even
no {\em a priori} knowledge of the future workload, while still incurring {\em
near-optimal} instance acquisition costs? To our knowledge, this paper represents
the first attempt to answer this question, as we make the following
contributions.

With dynamic programming, we first characterize the optimal offline reservation
strategy as a benchmark algorithm (Sec.~\ref{sec:offline}), in which the exact
future demand is assumed to be known {\em a priori}. We show that the optimal
strategy suffers ``the curse of dimensionality'' \cite{adp-powell} and is hence
computationally intractable. This indicates that optimal instance reservation
is in fact very difficult to obtain, even given the entire future demands.

Despite the complexity of the reservation problem in the offline setting, we
present two {\em online} reservation algorithms, one deterministic and another
randomized, that offer {\em the best provable} cost guarantees {\em without}
any knowledge of future demands beforehand. We first show that our
deterministic algorithm incurs no more than $2-\alpha$ times the minimum cost
obtained by the benchmark optimal offline algorithm (Sec.~\ref{sec:deter}), and
is therefore {\em $(2-\alpha)$-competitive}, where $\alpha \in [0,1]$ is the
entitled usage discount offered by reserved instances. This translates to a
worst-case cost that is 1.51 times the optimal one under the prevalent pricing
of Amazon EC2.  We then establish the more encouraging result that, our
randomized algorithm improves the competitive ratio to $e/(e-1+\alpha)$ in
expectation, and is 1.23-competitive under Amazon EC2 pricing
(Sec.~\ref{sec:rand}). Both algorithms achieve the {\em best possible}
competitive ratios in the deterministic and randomized cases, respectively, and
are simple enough for practical implementations. Our online algorithms can also
be extended to cases when short-term predictions into the near future are
reliable (Sec.~\ref{sec:prediction}).

In addition to our theoretical analysis, we have also evaluated both proposed
online algorithms via large-scale simulations (Sec.~\ref{sec:sim}), driven by
Google cluster-usage traces \cite{google-trace} with 40 GB workload demand
information of 933 users in one month. Our simulation results show that, under
the pricing of Amazon EC2 \cite{ec2pricing}, our algorithms closely track the
demand dynamics, realizing substantial cost savings compared with several
alternatives.

Though we focus on cost management of acquiring compute instances, our
algorithms may find wide applications in the prevalent IaaS services market.
For example, Amazon ElastiCache \cite{elasticache} also offers two pricing
options for its web caching services, i.e., the On-Demand Cache Nodes and
Reserved Cache Nodes, in which our proposed algorithms can be directly applied
to lower the service costs.


\section{Optimal Cost Management}
\label{sec:model}

We start off by briefly reviewing the pricing details of the on-demand and
reservation options in IaaS clouds, based on which we formulate the
online instance reservation problem for optimal cost management. 

\subsection{On-demand and Reservation Pricing}

{\bf On-Demand Instances:}
On-demand instances let users pay for compute capacity based on usage time without
long-term commitments, and are uniformly supported in leading IaaS clouds. For
example, in Amazon EC2, the hourly rate of a Standard Small Instance (Linux, US East) is
\$0.08 (see Table~\ref{tbl:ec2pricing}). In this case, running it on demand for
100 hours costs a user \$8.

On-demand instances resemble the conventional pay-as-you-go model.
Formally, for a certain type of instance, let the hourly rate be $p$.  Then
running it on demand for $h$ hours incurs a cost of $p h$. Note that in most
IaaS clouds, the hourly rate $p$ is set as fixed in a very long time period
(e.g., years), and can therefore be viewed as a constant. 

{\bf Reserved Instances:}
Another type of pricing option that is widely supported in IaaS clouds is the
reserved instance. It allows a user to reserve an instance for a long period
(months or years) by prepaying an upfront reservation fee, after which, the
usage is either free, e.g., ElasticHosts \cite{elastichosts}, GoGrid
\cite{gogrid}, or is priced with a heavy discount, e.g., Amazon EC2
\cite{ec2pricing}. For example, in Amazon EC2, to reserve a Standard Small Instance
(Linux, US East, Light Utilization) for 1 year, a user pays an upfront \$69 and
receives a discount rate of \$0.039 per hour within 1 year of the reservation
time, as oppose to the regular rate of \$0.08 (see Table~\ref{tbl:ec2pricing}).
Suppose this instance has run 100 hours before the reservation expires. Then
the total cost incurred is \$69 + 0.039$\times$100 = \$72.9.

Reserved instances resemble the wholesale market. Formally, for a certain
type of reserved instance, let the reservation period be $\tau$ (counted by the
number of hours). An instance that is reserved at hour $i$ would expire before
hour $i+\tau$. Without loss of generality, we assume the reservation fee to be
$1$ and normalize the on-demand rate $p$ to the reservation fee. Let $\alpha
\in [0,1]$ be the received discount due to reservation. A reserved instance
running for $h$ hours during the reservation period incurs a discounted running
cost $\alpha p h$ plus a reservation fee, leading to a total cost of $1 +
\alpha p h$. In the previous example, the normalized on-demand rate $p = 0.08 /
69$; the received discount due to reservation is $\alpha = 0.039 / 0.08 =
0.49$; the running hour $h = 100$; and the normalized overall cost is
\begin{equation*} 
  1 + \alpha p h = 72.9/69~.
\end{equation*}

In practice, cloud providers may offer multiple types of reserved instances
with different reservation periods and utilization levels. For example, Amazon
EC2 offers 1-year and 3-year reservations with light, medium, and high
utilizations \cite{ec2pricing}. For simplicity, we limit the discussion to one
type of such reserved instances chosen by a user based on its rough
estimations. We also assume that the on-demand rate is far smaller than the
reservation fee, i.e., $p \ll 1$, which is always the case in IaaS clouds,
e.g., \cite{ec2pricing, elastichosts, gogrid}. 

\subsection{The Online Instance Reservation Problem}




In general, launching instances on demand is more cost efficient for sporadic
workload, while reserved instances are more suitable to serve stable demand
lasting for a long period of time, for which the low hourly rate would
compensate for the high upfront fee. The cost management problem is to
optimally combine the two instance options to serve the time-varying demand,
such that the incurred cost is minimized. In this section, we consider making
instance purchase decisions {\em online}, without any {\em a priori} knowledge
about the future demands. Such an online model is especially important for
startup companies who have limited or no history demand data and those cloud
users whose workloads are highly variable and non-stationary --- in both cases
reliable predictions are unavailable. We postpone the discussions for cases
when short-term demand predictions are reliable in Sec.~\ref{sec:prediction}.

Since IaaS instances are billed in an hourly manner, we slot the time to a
sequence of hours indexed by $t=0,1,2,\dots$  At each time $t$, demand $d_t$
arrives, meaning that a user requests $d_t$ instances, $d_t = 0,1,2,\dots$ To
accommodate this demand, the user decides to use $o_t$ on-demand instances and
$d_t-o_t$ reserved instances.  If the previously reserved instances that remain
available at time $t$ are fewer than $d_t-o_t$, then new instances need to be
reserved.  Let $r_t$ be the number of instances that are {\em newly reserved}
at time $t$, $r_t = 0, 1, 2, \dots$ The overall cost incurred at time $t$ is
the on-demand cost $o_t p$ plus the reservation cost $r_t + \alpha p (d_t -
o_t)$, where $r_t$ is the upfront payments due to new reservations, and $\alpha
p (d_t - o_t)$ is the cost of running $d_t - o_t$ reserved instances. 



The cost management problem is to make instance purchase decisions online,
i.e., $r_t$ and $o_t$ at each time $t$, before seeing future demands $d_{t+1},
d_{t+2}, \dots$ The objective is to minimize the overall instance acquiring
costs. Suppose demands last for an arbitrary time $T$ (counted by the number of
hours). We have the following {\em online instance reservation} problem:
\begin{equation}
  \label{eq:opt-resv}
  \begin{split}
    \min_{\{r_t, o_t\}} & \quad C = \sum_{t=1}^{T}
    (o_t p + r_t + \alpha p (d_t - o_t))~, \\
    \mbox{s.t.} & \quad o_t + \sum_{i=t-\tau+1}^t r_i \geq d_t~, \\
    & \quad o_t, r_t \in \{0,1,2,\dots\}, t=1,\dots,T~.
  \end{split}
\end{equation}
Here, the first constraint ensures that all $d_t$ instances demanded at time
$t$ are accommodated, with $o_t$ on-demand instances and $\sum_{i=t-\tau+1}^t
r_i$ reserved instances that remain active at time $t$. Note that instances
that are reserved before time $t-\tau+1$ have all expired at time $t$, where
$\tau$ is the reservation period. For convenience, we set $r_t = 0$ for all $t
\leq 0$.

The main challenge of problem (\ref{eq:opt-resv}) lies in its online
setting. Without knowledge of future demands, the online strategy may
make purchase decisions that turn out later not to be optimal. Below we clarify
the performance metrics to measure how far away an online strategy may deviate
from the optimal solution.


\subsection{Measure of Competitiveness}

To measure the cost performance of an online strategy, we adopt the standard
{\em competitive analysis} \cite{borodin98}. The idea is to bound the gap
between the cost of an interested online algorithm and that of the optimal
offline strategy. The latter is obtained by solving problem \eqref{eq:opt-resv}
with the exact future demands $d_1, \dots, d_T$ given {\em a priori}. Formally, we
have

\begin{definition}[Competitive analysis]
  \label{def:comp-analy}
  A {\em deterministic} online reservation algorithm $A$ is {\em
  $c$-competitive} ($c$ is a constant) if for all possible demand sequences $\mathbf{d}
  = \{d_1, \dots, d_T\}$, we have
  \begin{equation}
    \label{eq:comp-analysis}
    C_A(\mathbf{d}) \le c \cdot C_\mathrm{OPT}(\mathbf{d})~,
  \end{equation}
  where $C_A(\mathbf{d})$ is the instance acquiring cost incurred by algorithm $A$ given input $\mathbf{d}$, and
  $C_\mathrm{OPT}(\mathbf{d})$ is the optimal instance acquiring cost given input $\mathbf{d}$. Here,
  $C_\mathrm{OPT}(\mathbf{d})$ is obtained by solving the instance reservation problem
  (\ref{eq:opt-resv}) {\em offline}, where the exact demand sequence $\mathbf{d}$ is
  assumed to know {\em a priori}.
\end{definition}

A similar definition of the competitive analysis also extends to the {\em
randomized} online algorithm $A$, where the decision making is drawn from a
random distribution. In this case, the LHS of (\ref{eq:comp-analysis}) is
simply replaced by $\E[C_A(\mathbf{d})]$, the expected cost of randomized
algorithm $A$ given input $\mathbf{d}$. (See \cite{borodin98} for a detailed
discussion.)

Competitive analysis takes an optimal offline algorithm as a benchmark to
measure the cost performance of an online strategy. Intuitively, the smaller
the competitive ratio $c$ is, the more closely the online algorithm $A$
approaches the optimal solution. Our objective is to design {\em optimal online
algorithms} with the smallest competitive ratio.

We note that the instance reservation problem (\ref{eq:opt-resv}) captures the
Bahncard problem \cite{bahncard} as a special case when a user demands no more
than one instance at a time, i.e., $d_t \le 1$ for all $t$. The Bahncard problem models
online ticket purchasing on the German Federal Railway, where one can opt to
buy a Bahncard (reserve an instance) and to receive a discount on all trips
within one year of the purchase date. It has been shown in \cite{bahncard, tcp-ack} that
the lower bound of the competitive ratio is $2 - \alpha$ and $e/(e-1+\alpha)$
for the deterministic and randomized Bahncard algorithms, respectively. Because the
Bahncard problem is a special case of our problem (\ref{eq:opt-resv}), we have

\begin{lemma}
  \label{lem:lower-bound}
  The competitive ratio of problem (\ref{eq:opt-resv}) is {\em at least} $2 -
  \alpha$ for deterministic online algorithms, and is at least
  $e/(e-1+\alpha)$ for randomized online algorithms.
\end{lemma}

However, we show in the following that the instance reserving problem
(\ref{eq:opt-resv}) is by no means a trivial extension to the Bahncard problem,
mainly due to the time-multiplexing nature of reserved instances.

\subsection{Bahncard Extension and Its Inefficiency}
\label{sec:bahncard}

A natural way to extend the Bahncard solutions in \cite{bahncard} is to
decompose problem (\ref{eq:opt-resv}) into separate Bahncard problems. To do
this, we introduce a set of {\em virtual users} indexed by 1, 2, \dots Whenever
demand $d_t$ arises at time $t$, we view the original user as $d_t$ virtual
users 1, 2, \dots, $d_t$, each requiring one instance at that time. Each virtual
user then reserves instances (i.e., buy a Bahncard) separately to minimize its
cost, which is exactly a Bahncard problem.

However, such an extension is highly inefficient. An instance reserved by one
virtual user, even idle, can {\em never} be multiplexed with another, who still
needs to pay for its own demand. For a real user, this implies that it has to
acquire additional instances, either on-demand or reserved, even if the user
has already reserved sufficient amount of instances to serve its demand, which
inevitably incurs a large amount of unnecessary cost.

We learn from the above failure that instances must be reserved {\em jointly}
and {\em time multiplexed} appropriately. These factors significantly
complicate our problem (\ref{eq:opt-resv}). Indeed, as we see in the next
section, even with full knowledge of the future demand, obtaining an optimal
offline solution to (\ref{eq:opt-resv}) is computationally prohibitive. 




\section{The Offline Strategy and Its Intractability}
\label{sec:offline}

In this section we consider the benchmark {\em offline cost management}
strategy for problem \eqref{eq:opt-resv}, in which the exact future demands are
given {\em a priori}. The offline setting is an integer programming problem and
is generally difficult to solve. We derive the optimal solution via dynamic
programming. However, such an optimal offline strategy suffers from ``the curse
of dimensionality'' \cite{adp-powell} and is computationally intractable.


We start by defining states. A state at time $t$ is defined as a
$(\tau-1)$-tuple  $\s_t = (s_{t,1}, \dots, s_{t,\tau - 1})$, where $s_{t,i}$ denotes the number of
instances that are reserved {\em no later than} $t$ and {\em remain active}
at time $t+i$, $i = 1, \dots, \tau-1$. We use a $(\tau-1)$-tuple to define a
state because an instance that is reserved no later than $t$ will no longer be
active at time $t+\tau$ and thereafter. Clearly, $s_{t,1} \ge \dots \ge
s_{t,\tau-1}$ as reservations gradually expire.

We make an important observation, that state $\s_t$ only depends on states
$\s_{t-1}$ at the previous time, and is {\em independent} of earlier states
$\s_{t-2}, \dots, \s_1$. Specifically, suppose state $\s_{t-1}$ is reached at
time $t-1$. At the beginning of the next time $t$, $r_t$ new instances are
reserved. These newly reserved $r_t$ instances will add to the active
reservations starting from time $t$, leading state $\s_{t-1}$ to transit to
$\s_t$ following the transition equations below:
\begin{equation}
  \label{eq:trans}
  \left\{
  \begin{array}{ll}
    s_{t,i} = s_{t-1,i+1} + r_t, & i = 1,\dots,\tau-2 ~; \\
    s_{t,\tau-1} = r_t. &
  \end{array}
  \right.
\end{equation}


Let $V(\s_t)$ be the minimum cost of serving demands $d_1,\dots,d_t$ up to
time $t$, conditioned upon the fact that state $\s_t$ is reached at time $t$. We
have the following recursive Bellman equations:
\begin{equation}
  \label{eq:bellman}
  V(\s_t) = \min_{s_{t-1}} \big\{ V(\s_{t-1}) + c(\s_{t-1}, \s_t) \big\}, \quad t>0,
\end{equation}
where $c(\s_{t-1}, \s_t)$ is the transition cost, and the minimization is over all states $\s_{t-1}$ that can transit to $\s_t$
following the transition equations (\ref{eq:trans}). The Bellman equations
(\ref{eq:bellman}) indicate that the minimum cost of reaching
$\s_t$ is given by the minimum cost of reaching a previous state $\s_{t-1}$
plus the transition cost $c(\s_{t-1}, \s_t)$, minimized over all possible
previous states $\s_{t-1}$. Let 
\begin{equation}
  X^+ = \max \{ 0, X \}~. 
\end{equation}
The transition cost is defined as
\begin{equation}
  c(\s_{t-1}, \s_t) = o_tp + r_t + \alpha p (d_t - o_t)~,
  \label{eq:trans-cost}
\end{equation}
where
\begin{equation}
  r_t = s_{t,\tau-1},
\end{equation}
\begin{equation}
  o_t = (d_t - r_t - s_{t-1,1})^+ ,
\end{equation}
and the transition from $\s_{t-1}$ to $\s_t$ follows (\ref{eq:trans}). 
The rationale of (\ref{eq:trans-cost}) is
straightforward. By the transition equations (\ref{eq:trans}), state $\s_{t-1}$
transits to $\s_t$ by reserving $r_t = s_{t,\tau-1}$ instances at time $t$.
Adding the $s_{t-1,1}$ instances that have been reserved before $t$, we have
$r_t + s_{t-1,1}$ reserved instances to use at time $t$. We therefore need $o_t =
(d_t - r_t - s_{t-1,1})^+$ on-demand instances at that time.

The boundary conditions of Bellman equations (\ref{eq:bellman}) are
\begin{equation}
  V(\s_0) = s_{0,1}, \quad \mbox{for all $\s_0 = (s_{0,1},\dots,s_{0,\tau-1})$},
  \label{eq:boundary}
\end{equation}
because an initial state $\s_0$ indicates that a user has already reserved
$s_{0,1}$ instances at the beginning and paid $s_{0,1}$.

With the analyses above, we see that the dynamic programming defined by
(\ref{eq:trans}), (\ref{eq:bellman}), (\ref{eq:trans-cost}), and
(\ref{eq:boundary}) optimally solves the offline instance reserving problem
(\ref{eq:opt-resv}). Therefore, it gives $C_{\mathrm{OPT}}(\mathbf{d})$ in
theory.


Unfortunately, the dynamic programming presented above is {\em computationally
intractable}. This is because to solve the Bellman equations
(\ref{eq:bellman}), one has to compute $V(\s_t)$ for all states $\s_t$.
However, since a state $\s_t$ is defined in a high-dimensional space --- recall
that $\s_t$ is defined as a $(\tau-1)$-tuple --- there exist {\em
exponentially} many such states. Therefore, looping over all of them results in
exponential time complexity.  This is known as the curse of dimensionality
suffered by high-dimensional dynamic programming \cite{adp-powell}.

The intractability of the offline instance reservation problem
\eqref{eq:opt-resv} suggests that optimal cost management in IaaS clouds is in
fact a very complicate problem, even if future demands can be accurately
predicted. However, we show in the following sections that it is possible to
have online strategies that are highly efficient with near-optimal cost
performance, even without any knowledge of the future demands.

\section{Optimal Deterministic Online Strategy}
\label{sec:deter}

In this section, we present a deterministic online reservation strategy that
incurs no more than $2-\alpha$ times the minimum cost. As indicated by
Lemma~\ref{lem:lower-bound}, this is also the best that one can expect from a
deterministic algorithm.

\subsection{The Deterministic Online Algorithm}



We start off by defining a {\em break-even point} at which a user is
indifferent between using a reserved instance and an on-demand instance.
Suppose an on-demand instance is used to accommodate workload in a time
interval that spans a reservation period, incurring a cost $c$. If we use a
reserved instance instead to serve the same demand, the cost will be $1 +
\alpha c$. When $c = 1/(1-\alpha)$, both instances cost the same, and are
therefore indifferent to the user. We hence define the break-even point as 
\begin{equation}
  \beta = 1/(1-\alpha)~. 
\end{equation}
Clearly, the use of an on-demand instance is well justified {\em if and only
if} the incurred cost does not exceed the break-even point, i.e.,
$c \le \beta$.

Our deterministic online algorithm is summarized as follows. By default, all
workloads are assumed to be operated with on-demand instances. At time $t$,
upon the arrival of demand $d_t$, we check the use of on-demand instances in a
recent reservation period, starting from time $t-\tau+1$ to $t$, and reserve a
new instance whenever we see an on-demand instance incurring more costs than
the break-even point. Algorithm~\ref{alg:deter} presents the detail.

\begin{algorithm}
  \caption{Deterministic Online Algorithm $A_\beta$}
  \label{alg:deter}
  \algsetup{linenosize=\footnotesize, linenodelimiter=.}
  \begin{algorithmic}[1]
    \STATE Let $x_i$ be the number of reserved instances
    at time $i$, Initially, $x_i \leftarrow 0$ for all $i = 0,1,\dots$
    \STATE Let $I(X)$ be an indicator function where $I(X) = 1$ if $X$ is true and $I(X) = 0$
    otherwise. Also let $X^+ = \max \{ X, 0 \}$.
    \STATE Upon the arrival of demand $d_t$, loop as follows:
    \label{lne:dt}
    \WHILE {$p \sum_{i=t-\tau+1}^t I(d_i > x_i) > \beta$}
    \label{lne:while}
	\STATE Reserve a new instance: $r_t \leftarrow r_t + 1$.
	\label{lne:reserv}
	\STATE Update the number of reservations that can be used in the future:
	$x_i \leftarrow x_i + 1$ for $i = t, \dots, t+\tau-1$.
	\label{lne:future}
	\STATE Add a ``phantom'' reservation to the recent period, indicating that the history has
	already been ``processed'': $x_i \leftarrow x_i + 1$ for $i = t-\tau+1, \dots, t-1$.
	\label{lne:phantom}
    \ENDWHILE
    \STATE Launch on-demand instances: $o_t \leftarrow (d_t - x_t)^+$.
    \STATE $t \leftarrow t + 1$, repeat from \ref{lne:dt}.
  \end{algorithmic}
\end{algorithm}

\begin{figure}[tb]
  \centering
  \includegraphics[width=0.43\textwidth]{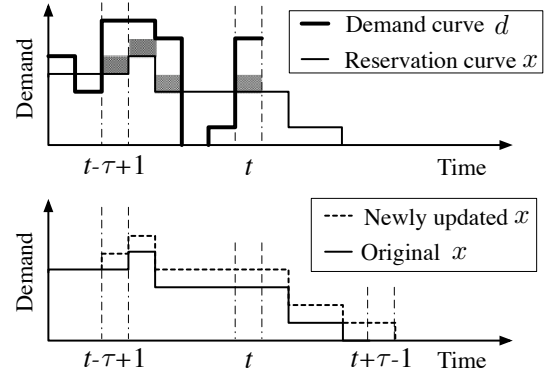}
  \vspace{-.1in}
  \caption{Illustration of Algorithm~\ref{alg:deter}. The shaded area in the
  top figure shows the use of an on-demand instance in the recent period. An instance
  is reserved at time $t$ if the use of this on-demand instance is not well justified.
  The bottom figure shows the corresponding updates of the reservation curve $x$.
  }
  \vspace{-.1in}
  \label{fig:deter}
\end{figure}

Fig.~\ref{fig:deter} helps to illustrate Algorithm~\ref{alg:deter}. Whenever
demand $d_t$ arises, we check the recent reservation period from time $t-\tau+1$ to
$t$. We see that an on-demand instance has been used at time $i$ if demand
$d_i$ exceeds the number of reservations $x_i$ (both actual and phantom), $i
= t-\tau+1, \dots, t$. The shaded area in Fig.~\ref{fig:deter} represents the
use of an on-demand instance in the recent period, which incurs a cost of $p
\sum_{i=t-\tau+1}^{t} I(d_i > x_i)$. If this cost exceeds the break-even point $\beta$
(line~\ref{lne:while} of Algorithm~\ref{alg:deter}), then such use of an on-demand instance
is not well justified: We {\em should have} reserved an instance before
at time $t-\tau+1$ and used it to serve the demand (shaded area) instead, which {\em would
have} lowered the cost. As a compensation for this ``mistake,'' we reserve an
instance at the current time $t$ (line~\ref{lne:reserv}), and will have one
more reservation to use in the future (line~\ref{lne:future}). Since we have
already compensated for a misuse of an on-demand instance (the shaded area), we
add a ``phantom'' reservation to the history so that such a mistake will not be
counted multiple times in the following rounds (line~\ref{lne:phantom}). This leads
to an update of the reservation number $\{x_i\}$ (see the bottom figure in
Fig.~\ref{fig:deter}).

Unlike the simple extension of the Bahncard algorithm described in
Sec.~\ref{sec:bahncard}, Algorithm~\ref{alg:deter} jointly reserves instances
by taking both the currently active reservations (i.e., $x_t$) and the historic
records (i.e., $x_i$, $i < t$) into consideration (line~\ref{lne:while}),
without any knowledge of the future. We will see later in Sec.~\ref{sec:sim}
that such a joint reservation significantly outperforms the Bahncard extension
where instances are reserved separately.

\subsection{Performance Analysis: $(2-\alpha)$-Competitiveness}

The ``trick'' of Algorithm~\ref{alg:deter} is to make reservations ``lazily'': no
instance is reserved unless the misuse of an on-demand instance is seen. Such 
a ``lazy behaviour'' turns out to guarantee that the algorithm incurs no more than $2-\alpha$
times the minimum cost.

Let $A_{\beta}$ denote Algorithm~\ref{alg:deter} and let OPT denote the optimal
offline algorithm. We now make an important observation, that OPT reserves at least
the same amount of instances as $A_\beta$ does, for any demand sequence.

\begin{lemma}
  \label{lem:resv-num}
  Given an arbitrary demand sequence, let $n_\beta$ be the number of instances
  reserved by $A_\beta$, and let $n_{\mathrm{OPT}}$ be the number of instances reserved
  by OPT. Then $n_\beta \le n_{\mathrm{OPT}}$.
\end{lemma}

Lemma~\ref{lem:resv-num} can be viewed as a result of the ``lazy behaviour'' of
$A_\beta$, in which instances are reserved just to compensate for the previous
``purchase mistakes.'' Intuitively, such a conservative reservation strategy
leads to fewer reserved instances. The proof of Lemma~\ref{lem:resv-num} is
is given in Appendix~\ref{app:deter}.

We are now ready to analyze the cost performance of $A_\beta$, using the
optimal offline algorithm OPT as a benchmark.

\begin{proposition}
  \label{prop:deter-comp}
  Algorithm~\ref{alg:deter} is {\em $(2-\alpha)$-competitive}. Formally, for
  any demand sequence,
  \begin{equation}
    C_{A_\beta} \le (2-\alpha) C_{\mathrm{OPT}}~,
  \end{equation}
  where $C_{A_\beta}$ is the cost of Algorithm~\ref{alg:deter} ($A_\beta$), and $C_{\mathrm{OPT}}$ is
  the cost of the optimal offline algorithm OPT.
\end{proposition}

{\bf Proof:}
  Suppose $A_\beta$ (resp., OPT) launches $o_t$ (resp., $o_t^*$) on-demand
  instances at time $t$.  Let $\mathrm{Od}(A_\beta)$ be
  the costs incurred by these on-demand instances under $A_\beta$, i.e.,
  \begin{equation}
    \mathrm{Od}(A_\beta) = \sum_{t=1}^T o_t p ~.
  \end{equation}
  We refer to $\mathrm{Od}(A_\beta)$ as the {\em on-demand costs} of $A_\beta$.
  Similarly, we define the on-demand costs incurred by OPT as
  \begin{equation}
    \mathrm{Od}(\mathrm{OPT}) = \sum_{t=1}^T o_t^* p~. 
  \end{equation}  
  Also, let 
  \begin{equation}
    \mathrm{Od}(A_\beta \backslash \mathrm{OPT}) = \sum_{t=1}^T (o_t - o_t^*)^+ p 
  \end{equation}
  be the on-demand costs incurred in $A_\beta$ that are not incurred in OPT. We see
  \begin{equation}
    \label{eq:od-cost}
    \mathrm{Od}(A_\beta \backslash \mathrm{OPT}) \le \beta n_\mathrm{OPT}
  \end{equation}
  by noting the following two facts: First, demands $\sum_{t=1}^T (o_t - o_t^*)^+$
  are served by at most $n_\mathrm{OPT}$ reserved instances in OPT. Second,
  demands that are served by the same reserved instance in OPT incur on-demand costs
  of at most $\beta$ in $A_\beta$ (by the definition of $A_\beta$). We
  therefore bound $\mathrm{Od}(A_\beta)$ as follows:
  \begin{align}
    \mathrm{Od}(A_\beta) & \le
    \mathrm{Od}(\mathrm{OPT}) + \mathrm{Od}(A_\beta \backslash \mathrm{OPT}) \nonumber \\
    & \le \mathrm{Od}(\mathrm{OPT}) + \beta n_\mathrm{OPT}~.
    \label{eq:reg-cost}
  \end{align}



  Let $S=\sum_{t=1}^T d_t p$ be the cost of serving all demands with on-demand instances.
  We bound the cost of OPT as follows:
  \begin{align}
    C_\mathrm{OPT} & = \mathrm{Od}(\mathrm{OPT}) + n_\mathrm{OPT} + \alpha
    (S - \mathrm{Od}(\mathrm{OPT})) \\
    & \ge \mathrm{Od}(\mathrm{OPT}) + n_\mathrm{OPT} + \alpha \beta n_\mathrm{OPT}
    \label{eq:opt-1-neq} \\
    & \ge n_\mathrm{OPT} / (1-\alpha) ~.
    \label{eq:opt-bound}
  \end{align}
  Here, (\ref{eq:opt-1-neq}) holds because in OPT, demands that are served by
  the same reserved instance incur at least a break-even cost $\beta$ when
  priced at an on-demand rate $p$.

  With (\ref{eq:reg-cost}) and (\ref{eq:opt-bound}), we bound the cost of $A_\beta$ as follows:
  \begin{align}
    C_{A_\beta} & = \mathrm{Od}(A_\beta) + n_\beta + \alpha
    (S - \mathrm{Od}(A_\beta)) \nonumber \\
    & \le (1-\alpha) \mathrm{Od}(A_\beta) + n_\mathrm{OPT} + \alpha S
    \label{eq:deter-1-neq} \\
    & \le (1-\alpha) (\mathrm{Od}(\mathrm{OPT}) + \beta n_\mathrm{OPT} )
    + \alpha S + n_\mathrm{OPT}
    \label{eq:deter-2-neq} \\
    & = C_\mathrm{OPT} + n_\mathrm{OPT} \label{eq:deter-2-eq} \\
    & \le (2-\alpha) C_\mathrm{OPT} ~.
    \label{eq:deter-bound}
  \end{align}
  Here, (\ref{eq:deter-1-neq}) holds because $n_\beta \le n_\mathrm{OPT}$
  (Lemma~\ref{lem:resv-num}). Inequality (\ref{eq:deter-2-neq}) follows
  from (\ref{eq:reg-cost}), while (\ref{eq:deter-bound}) is derived from (\ref{eq:opt-bound}).
\qed

By Lemma~\ref{lem:lower-bound}, we see that $2-\alpha$ is already the best
possible competitive ratio for deterministic online algorithms, which implies
that Algorithm~\ref{alg:deter} is optimal in a view of competitive analysis.

\begin{proposition}
  Among all online deterministic algorithms of problem (\ref{eq:opt-resv}),
  Algorithm~\ref{alg:deter} is {\em optimal} with the smallest competitive ratio of $2-\alpha$.
\end{proposition}

As a direct application, in Amazon EC2 with reservation discount $\alpha =
0.49$ (see Table~\ref{tbl:ec2pricing}), algorithm $A_\beta$ will lead to no more than 1.51
times the optimal instance purchase cost.

Despite the already satisfactory cost performance offered by the proposed
deterministic algorithm, we show in the next section that the competitive ratio
may be further improved if randomness is introduced.

\section{Optimal Randomized Online Strategy}
\label{sec:rand}

In this section, we construct a randomized online strategy that is a random
distribution over a family of deterministic online algorithms similar to
$A_\beta$. We show that such randomization improves the competitive ratio to
$e/(e-1+\alpha)$ and hence leads to a better cost performance. As indicated by
Lemma~\ref{lem:lower-bound}, this is the best that one can expect without
knowledge of future demands.

\subsection{The Randomized Online Algorithm}

We start by defining a family of algorithms similar to the deterministic
algorithm $A_\beta$. Let $A_z$ be a similar deterministic algorithm to
$A_\beta$ with $\beta$ in line~\ref{lne:while} of Algorithm~\ref{alg:deter}
replaced by $z \in [0, \beta]$. That is, $A_z$ reserves an instance whenever it
sees an on-demand instance incurring more costs than $z$ in the recent
reservation period.  Intuitively, the value of $z$ reflects the {\em
aggressiveness} of a reservation strategy. The smaller the $z$, the more
aggressive the strategy. As an extreme, a user will always reserve when $z =
0$. Another extreme goes to $z=\beta$ (Algorithm~\ref{alg:deter}), in which the
user is very conservative in reserving new instances.

Our randomized online algorithm picks a $z \in [0, \beta]$ according to a
density function $f(z)$ and runs the resulting algorithm $A_z$. Specifically,
the density function $f(z)$ is defined as
\begin{equation}
  \label{eq:density}
  f(z) = \left\{
  	\begin{array}{ll}
	  (1-\alpha)e^{(1-\alpha)z}/(e-1+\alpha), & z \in [0,\beta), \\
	  \delta(z - \beta) \cdot \alpha / (e-1+\alpha), & \mbox{o.w.,}
	\end{array}
	\right.
\end{equation}
where $\delta(\cdot)$ is the {\em Dirac delta function}.
That is, we pick $z=\beta$ with probability $\alpha / (e-1+\alpha)$. It is
interesting to point out that in other online rent-or-buy problems, e.g.,
\cite{ski-rental, tcp-ack, Lu12b}, the density function of a randomized
algorithm is usually continuous\footnote{The density function in these works is
chosen as $f(z) = e^z/(e-1), z\in[0,1]$, which is a special case of ours when $\alpha =
0$.}. However, we note that a continuous density function does not lead to the
minimum competitive ratio in our problem.  Algorithm~\ref{alg:rand} formalizes
the descriptions above.

\begin{algorithm}
  \caption{Randomized Online Algorithm}
  \label{alg:rand}
  \algsetup{linenosize=\footnotesize, linenodelimiter=.}
  \begin{algorithmic}[1]
    \STATE Randomly pick $z \in [0, \beta]$ according to a density function $f(z)$ defined by (\ref{eq:density})
    \STATE Run $A_z$
  \end{algorithmic}
\end{algorithm}

The rationale behind Algorithm~\ref{alg:rand} is to strike a suitable balance
between reserving ``aggressively'' and ``conservatively.'' Intuitively, being
aggressive is cost efficient when future demands are long-lasting and stable,
while being conservative is efficient for sporadic demands. Given the unknown
future, the algorithm randomly chooses a strategy $A_z$, with an expectation
that the incurred cost will closely approach the {\em ex post} minimum cost. We
see in the following that the choice of $f(z)$ in (\ref{eq:density}) leads to
the optimal competitive ratio $e/(e-1+\alpha)$.

\subsection{Performance Analysis: $e/(e-1+\alpha)$-Competitiveness}

To analyze the cost performance of the randomized algorithm, we need to
understand how the cost of algorithm $A_z$ relates to the cost of the optimal
offline algorithm OPT. The following lemma reveals their relationship.  The
proof is given in Appendix~\ref{app:rand}.

\begin{lemma}
  \label{lem:rand}
  Given an arbitrary demand sequence $d_1,\dots,d_T$, suppose algorithm $A_z$ (resp., OPT) launches
  $o_{z,t}$ (resp., $o_t^*$) on-demand instances at time $t$. Let $C_{A_z}$ be the instance
  acquiring cost incurred by algorithm $A_z$, and $n_z$ the number of instances
  reserved by $A_z$. Denote by 
  \begin{equation}
    D_z = \sum_{t=1}^T (o_t^* - o_{z,t})^+ p
  \end{equation}
  the on-demand costs incurred by OPT that are not by algorithm $A_z$. We have
  the following three statements.

  (1) The cost of algorithm $A_z$ is at most
      \begin{equation}
	\label{eq:cost-az}
	C_{A_z} \le C_\mathrm{OPT} - n_\mathrm{OPT} + n_z + (1-\alpha)( z n_\mathrm{OPT} - D_z).
      \end{equation}

  (2) The value of $D_z$ is at least
      \begin{equation}
	\label{eq:ez}
	D_z \ge \int_z^{\beta} n_w \ud w - (\beta - z) n_\mathrm{OPT} ~.
      \end{equation}

  (3) The cost incurred by OPT is at least
      \begin{equation}
	\label{eq:cost-opt}
	C_\mathrm{OPT} \ge \int_0^{\beta} n_z \ud z ~.
      \end{equation}
\end{lemma}

With Lemma~\ref{lem:rand}, we bound the expected cost of our randomized
algorithm with respect to the cost incurred by OPT.

\begin{proposition}
  \label{prop:rand-comp}
  Algorithm~\ref{alg:rand} is $e/(e-1+\alpha)$-competitive. Formally, for any demand sequence,
  \begin{equation}
    \E [ C_{A_z} ] \le \frac{e}{e-1+\alpha} C_{\mathrm{OPT}}~,
  \end{equation}
  where the expectation is over $z$ between 0 and $\beta$ according to density function
  $f(z)$ defined in (\ref{eq:density}).
\end{proposition}

{\bf Proof:}
  Let $F(z) = \int_0^z f(x) \ud x$, and $E_F = \int_0^z x f(x) \ud x$. From \eqref{eq:cost-az}, we have
  \begin{align}
    \E[C_{A_z}] & \le C_\mathrm{OPT} - n_\mathrm{OPT} + \mbox{$\int_0^\beta f(z) n_z \ud z$} \nonumber \\
    & \quad + (1-\alpha) \mbox{$\int_0^\beta f(z)(z  n_\mathrm{OPT} - D_z) \ud z$}
    \label{eq:rand-1-neq} \\
    & \le C_\mathrm{OPT} - \alpha n_\mathrm{OPT} E_F + \mbox{$\int_0^\beta f(z) n_z \ud z$} \nonumber \\
    & \quad - (1-\alpha) \mbox{$\int_0^\beta n_w \int_0^w f(z) \ud z \ud w$}
    \label{eq:rand-2-neq} \\
    & = C_\mathrm{OPT} + \mbox{$\int_0^\beta ( f(z) - (1-\alpha)F(z) ) n_z \ud z$} \nonumber \\
    & \quad - \alpha n_\mathrm{OPT} E_F ~,
    \label{eq:rand-3-neq}
  \end{align}
  where the second inequality is obtained by plugging in inequality (\ref{eq:ez}).

  Now divide both sides of inequality (\ref{eq:rand-3-neq}) by
  $C_\mathrm{OPT}$ and apply inequality (\ref{eq:cost-opt}). We have
  \begin{equation}
    \label{eq:comp-ratio}
    \frac{\E[C_{A_z}]}{C_\mathrm{OPT} } \le 1 +
    \frac{ \int_0^\beta ( f(z) - (1-\alpha)F(z) ) n_z \ud z  -  \alpha n_\mathrm{OPT} E_F}{ \int_0^\beta n_z
    \ud z }~.
  \end{equation}
  Plugging $f(z)$ defined in (\ref{eq:density}) into (\ref{eq:comp-ratio}) and
  noting that $n_\beta \le n_\mathrm{OPT}$ (Lemma~\ref{lem:resv-num}) lead to
  the desired competitive ratio.
\qed

By Lemma~\ref{lem:lower-bound}, we see that no online randomized algorithm is
better than Algorithm~\ref{alg:rand} in terms of the competitive ratio.

\begin{proposition}
  Among all online randomized algorithms of problem (\ref{eq:opt-resv}),
  Algorithm~\ref{alg:rand} is optimal with the smallest competitive ratio $e/(e-1+\alpha)$.
\end{proposition}

\begin{figure}[tb]
  \centering
  \includegraphics[width=0.33\textwidth]{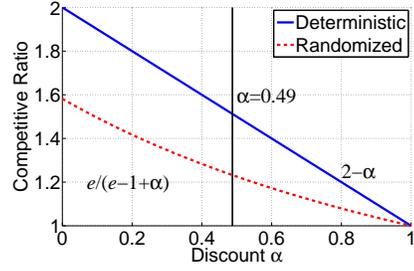}
  \caption{Competitive ratios of both deterministic and randomized algorithms.}
  \label{fig:comp-ratio}
  \vspace{-.1in}
\end{figure}

We visualize in Fig.~\ref{fig:comp-ratio} the competitive ratios of both
deterministic and randomized algorithms against the hourly discount $\alpha$
offered by reserved instances. Compared with the deterministic algorithm, we
see that introducing randomness significantly improves the competitive ratio in
all cases. The two algorithms become exactly the same when $\alpha = 1$, at
which the reservation offers no discount and will never be considered. In
particular, when it comes to Amazon EC2 with reservation discount $\alpha =
0.49$ (standard 1-year reservation, light utilization), the randomized
algorithm leads to a competitive ratio of 1.23, compared with the
1.51-competitiveness of the deterministic alternative. Yet, this does not imply
that the former always incurs less instance acquiring costs --- it is more
efficient than the deterministic alternative only in expectation.

\section{Cost Management with Short-Term Demand Predictions}
\label{sec:prediction}

In the previous sections, our discussions focus on the extreme cases, with
either full future demand information (i.e., the offline case in
Sec.~\ref{sec:offline}) or no {\em a priori} knowledge of the future (i.e., the
online case in Sec.~\ref{sec:deter} and \ref{sec:rand}). In this section, we
consider the middle ground in which short-term demand predictions are reliable.
For example, websites typically see diurnal patterns exhibited on their
workloads, based on which it is possible to have a demand prediction window
that is weeks into the future. Both our online algorithms can be easily
extended to utilize these knowledge of future demands when making reservation
decisions.

We begin by formulating the instance reservation problem with limited
information of future demands. Let $w$ be the prediction window. That is, at
any time $t$, a user can predict its future demands $d_{t+1}, \dots, d_{t+w}$
in the next $w$ hours. Since only short-term predictions are reliable, one can
safely assume that the prediction window is less than a reservation period,
i.e., $w < \tau$.  The instance reservation problem resembles the online
reservation problem \eqref{eq:opt-resv}, except that the instance purchase
decisions made at each time $t$, i.e., the number of reserved instances ($r_t$)
and on-demand instances ($o_t$), are based on both history and future demands
predicted, i.e., $d_1, \dots, d_{t+w}$. The competitive analysis
(Definition~\ref{def:comp-analy}) remains valid in this case.

{\bf The Deterministic Algorithm:} We extend our deterministic online algorithm
as follows. As before, all workloads are {\em by default} served by on-demand
instances. At time $t$, we can predict the demands up to time $t+w$. Unlike the
online deterministic algorithm, we check the use of on-demand instances in a reservation
period across {\em both history and future}, starting from time $t+w-\tau+1$ to
$t+w$. A new instance is reserved at time $t$ whenever we see an on-demand
instance incurring more costs than the break-even point $\beta$ and the
currently effective reservations are less than the current demand $d_t$.
Algorithm~\ref{alg:deter-pred}, also denoted by $A_\beta^w$, shows the details.

\begin{algorithm}
  \caption{Deterministic Algorithm $A_\beta^w$ with Prediction Window $w$}
  \label{alg:deter-pred}
  \algsetup{linenosize=\footnotesize, linenodelimiter=.}
  \begin{algorithmic}[1]
    \STATE Let $x_i$ be the number of reserved instances
    at time $i$, Initially, $x_i \leftarrow 0$ for all $i = 0,1,\dots$
    \STATE Upon the arrival of demand $d_t$, loop as follows:
    \label{lne:dt-pred}
    \WHILE {$p \sum_{i=t+w-\tau+1}^{t+w} I(d_i > x_i) > \beta$ and $x_t < d_t$}
    \label{lne:while-pred}
	\STATE Reserve a new instance: $r_t \leftarrow r_t + 1$.
	\label{lne:reserv-pred}
	\STATE Update the number of reservations that can be used in the future:
	$x_i \leftarrow x_i + 1$ for $i = t, \dots, t+\tau-1$.
	\label{lne:future-pred}
	\STATE Add a ``phantom'' reservation to the history, indicating that the history has
	already been ``processed'': $x_i \leftarrow x_i + 1$ for $i = t+w-\tau+1, \dots, t-1$.
	\label{lne:phantom-pred}
    \ENDWHILE
    \STATE Launch on-demand instances: $o_t \leftarrow (d_t - x_t)^+$.
    \STATE $t \leftarrow t + 1$, repeat from \ref{lne:dt-pred}.
  \end{algorithmic}
\end{algorithm}

{\bf The Randomized Algorithm:} 
The randomized algorithm can also be constructed as a random distribution over
a family of deterministic algorithms similar to $A_\beta^w$. In particular, let
$A_z^w$ be similarly defined as algorithm $A_\beta^w$ with $\beta$ replaced by
$z \in [0, \beta]$ in line~\ref{lne:while-pred} of
Algorithm~\ref{alg:deter-pred}. The value of $z$ reflects the aggressiveness of
instance reservation. The smaller the $z$, the more aggressive the reservation
strategy. Similar to the online randomized, we introduce randomness to strike a
good balance between reserving aggressively and conservatively. Our algorithm
randomly picks $z \in [0, \beta]$ according to the same density function $f(z)$
defined by \eqref{eq:density}, and runs the resulting algorithm $A_z^w$.
Algorithm~\ref{alg:rand-pred} formalizes the description above.

\begin{algorithm}
  \caption{Randomized Algorithm with Prediction Window $w$}
  \label{alg:rand-pred}
  \algsetup{linenosize=\footnotesize, linenodelimiter=.}
  \begin{algorithmic}[1]
    \STATE Randomly pick $z \in [0, \beta]$ according to a density function $f(z)$ defined by (\ref{eq:density})
    \STATE Run $A_z^w$
  \end{algorithmic}
\end{algorithm}

It is easy to see that both the deterministic and the randomized algorithms
presented above improve the cost performance of their online counterparts, due
to the knowledge of future demands. Therefore, we have
Proposition~\ref{prop:comp-pred} below.  We will quantify their performance
gains via trace-driven simulations in the next section. 

\begin{proposition}
  \label{prop:comp-pred}
  Algorithm~\ref{alg:deter-pred} is $(2-\alpha)$-competitive, and
  Algorithm~\ref{alg:rand-pred} is $e/(e-1+\alpha)$-competitive.
\end{proposition}

\section{Trace-Driven Simulations}
\label{sec:sim}

So far, we have analyzed the cost performance of the proposed algorithms in a
view of competitive analysis. In this section, we evaluate their performance
for practical cloud users via simulations driven by a large volume of
real-world traces.

\subsection{Dataset Description and Preprocessing}

Long-term user demand data in public IaaS clouds are often confidential: no
cloud provider has released such information so far. For this reason, we turn
to Google cluster-usage traces that were recently released in
\cite{google-trace}. Although Google is not a public IaaS cloud, its
cluster-usage traces record the computing demands of its cloud services and
Google engineers, which can represent the computing demands of IaaS users to
some degree. The dataset contains 40 GB of workload resource requirements
(e.g., CPU, memory, disk, etc.) of 933 users over 29 days in May 2011, on a
cluster of more than 11K Google machines.

\begin{figure}[tb]
  \centering
  \includegraphics[width=0.48\textwidth]{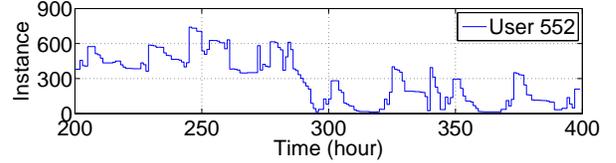}
  \vspace{-.1in}
  \caption{The demand curve of User 552 in Google cluster-usage
  traces \cite{google-trace}, over 1 month.
  }
  \vspace{-.1in}
  \label{fig:demand-curves}
\end{figure}

{\bf Demand Curve:} Given the workload traces of each user, we ask the question:
How many computing instances would this user require if it were to run the same
workload in a public IaaS cloud? For simplicity, we set an instance to have the
same computing capacity as a cluster machine, which enables us to accurately
estimate the run time of computational tasks by learning from the original traces.  We then
schedule these tasks onto instances with sufficient resources to accommodate
their requirements. Computational tasks that cannot run on the same server in
the traces (e.g., tasks of MapReduce) are scheduled to different instances. In
the end, we obtain a demand curve for each user, indicating how many instances
this user requires in each hour.  Fig.~\ref{fig:demand-curves} illustrates such
a demand curve for a user.

\begin{figure}[t]
  \centering
  \includegraphics[width=0.66\linewidth]{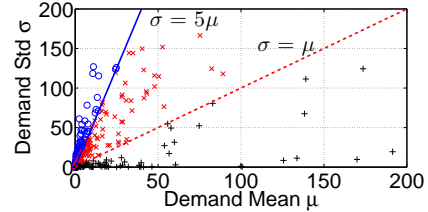}
  \caption{User demand statistics and group division.}
  \label{fig:stats}
  \vspace{-.1in}
\end{figure}

\begin{figure*}[t]
  \subfloat[Cost CDF (all users)]
  {
      \label{fig:all-fluc}
      \includegraphics[width=0.24\linewidth]{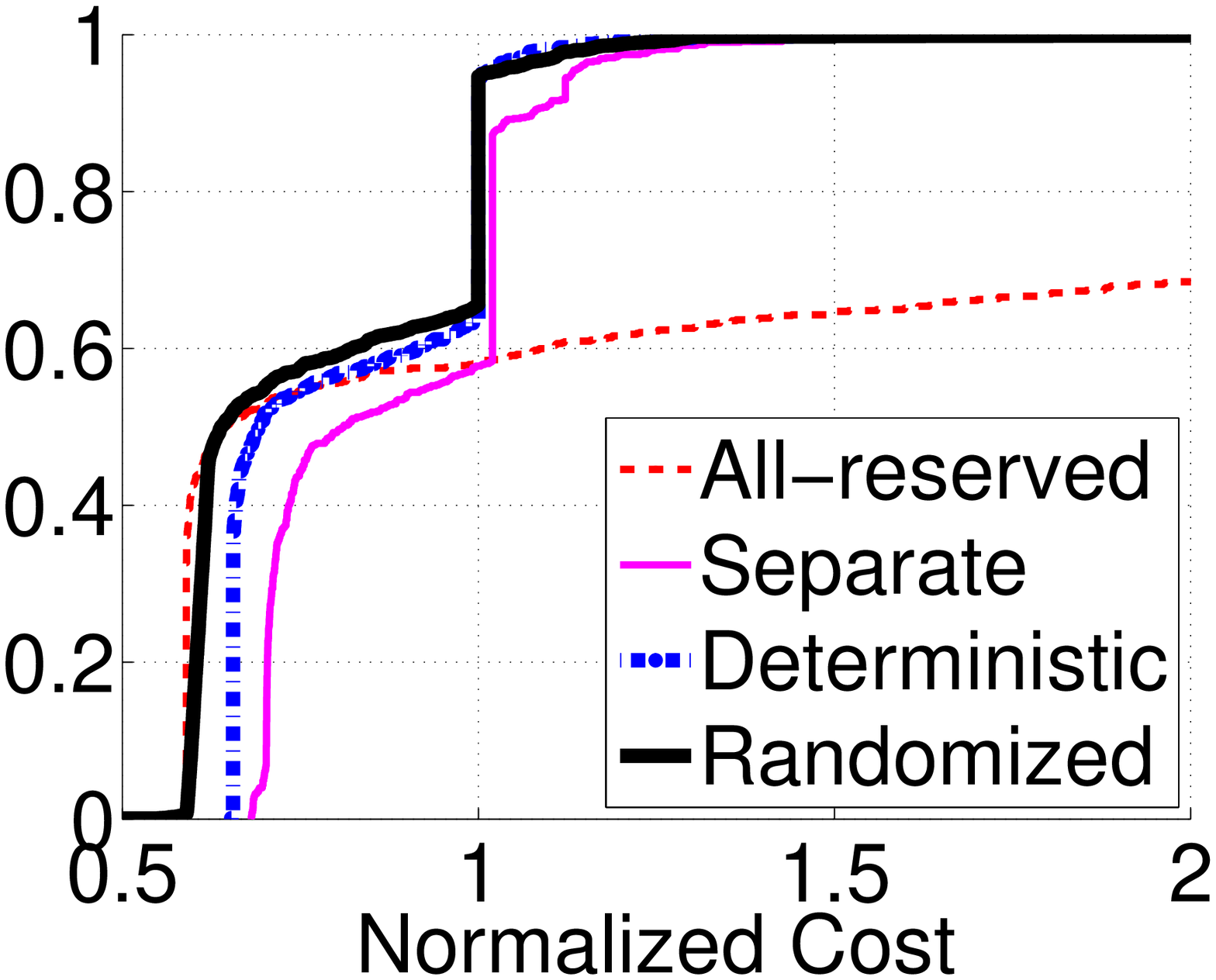}
  }
  \hspace{-2mm}
  \subfloat[Cost CDF (high fluctuation)]
  {
      \label{fig:high-fluc}
      \includegraphics[width=0.24\linewidth]{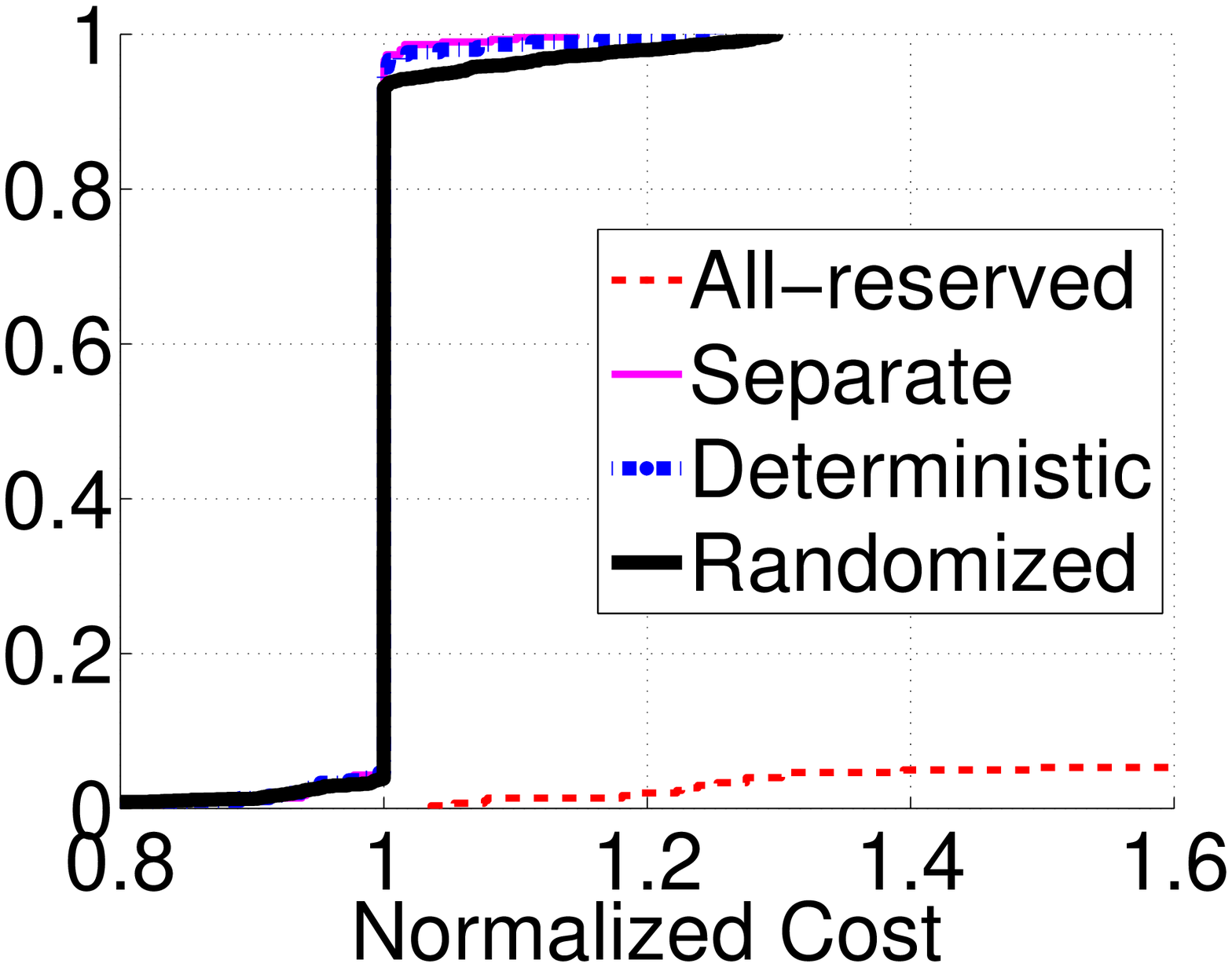}
  }
  \hspace{-2mm}
  \subfloat[Cost CDF (medium fluctuation)]
  {
      \label{fig:med-fluc}
      \includegraphics[width=0.24\linewidth]{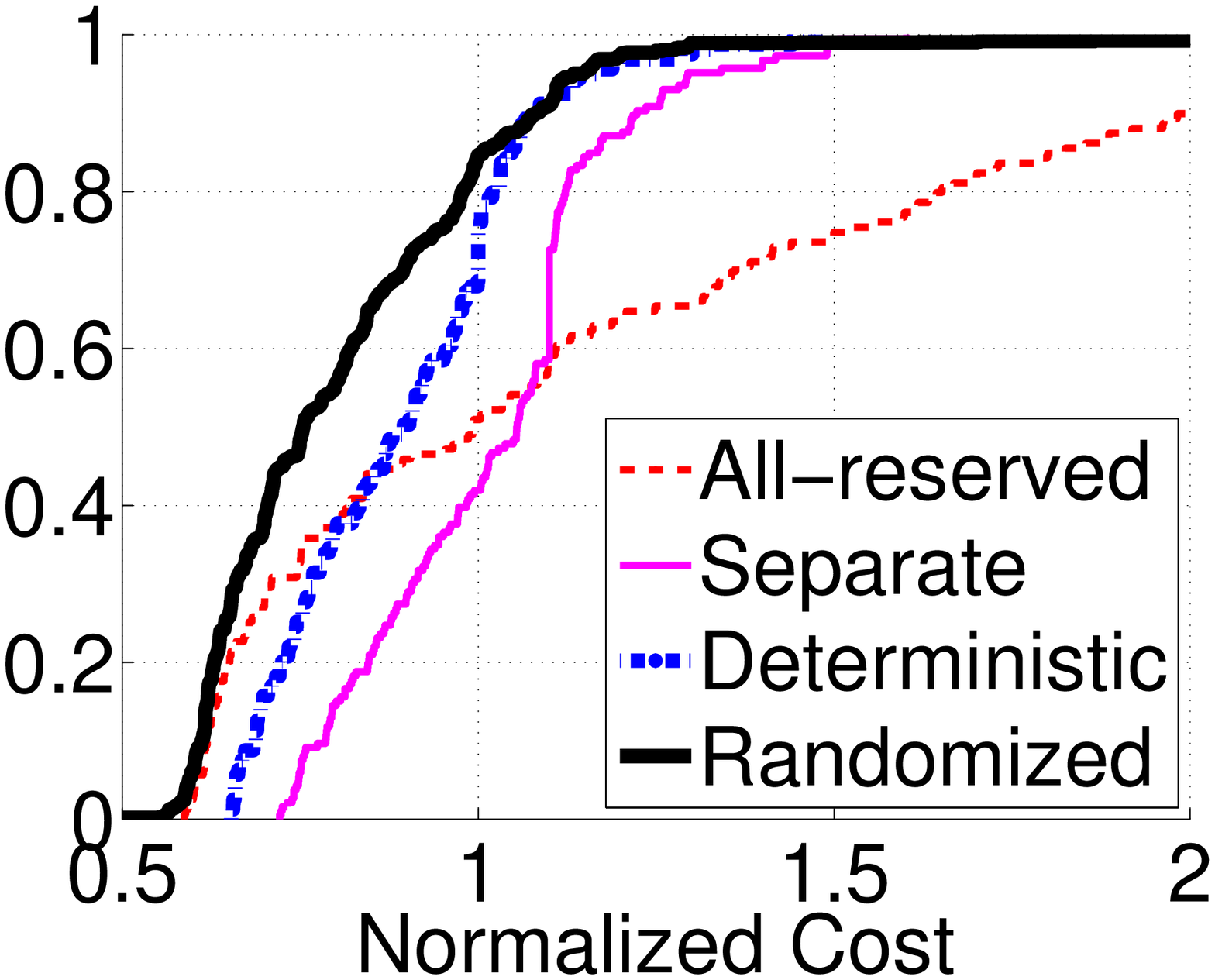}
  }
  \hspace{-2mm}
  \subfloat[Cost CDF (stable demands)]
  {
      \label{fig:low-fluc}
      \includegraphics[width=0.24\linewidth]{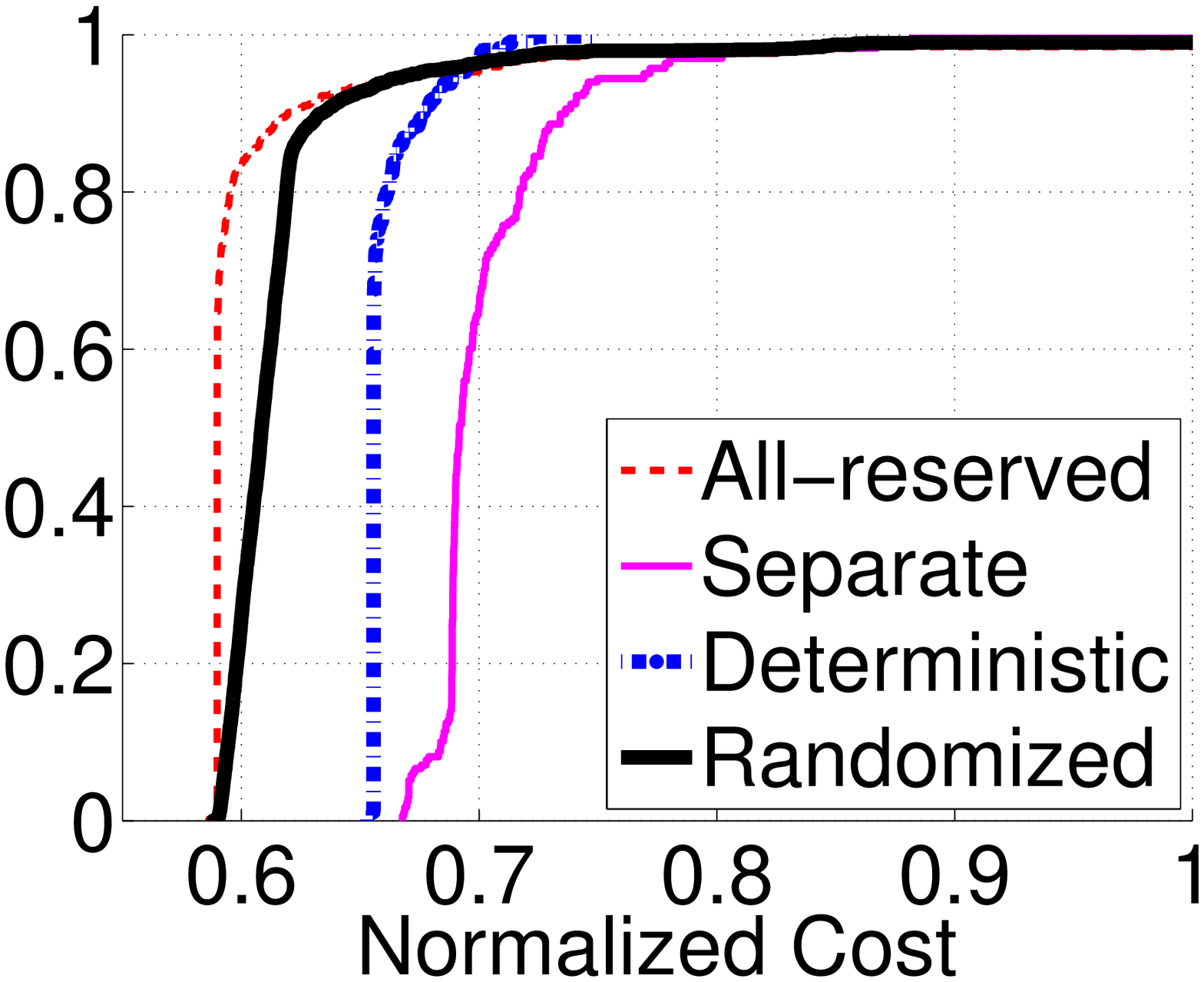}
  }
  \caption{
  Cost performance of online algorithms without {\em a priori} knowledge of future demands. 
  All costs are normalized to All-on-demand.
  }
  \label{fig:cost}
  \vspace{-2mm}			
\end{figure*}

{\bf User Classification:} To investigate how our online algorithms perform
under different demand patterns, we classify all 933 users into three groups by
the {\em demand fluctuation level} measured as the ratio between the
standard deviation $\sigma$ and the mean $\mu$.

Specifically, {\em Group 1} consists of users whose demands are highly
fluctuating, with $\sigma/\mu \ge 5$. As shown in Fig.~\ref{fig:stats} (circle `o'), these
demands usually have small means, which implies that they are highly sporadic
and are best served with on-demand instances. {\em Group 2} includes users
whose demands are less fluctuating, with $1 \le \sigma/\mu < 5$. As shown in
Fig.~\ref{fig:stats} (cross `x'), these demands cannot be simply served by
on-demand or reserved instances alone. {\em Group 3} includes all remaining
users with relatively stable demands ($0 \le \sigma/\mu < 1$). As shown in
Fig.~\ref{fig:stats} (plus `+'), these demands have large means and are best served
with reserved instances. Our evaluations are carried out for each user group.

{\bf Pricing:} Throughout the simulation, we adopt the pricing of Amazon EC2
standard small instances with the on-demand rate \$0.08, the reservation fee
\$69, and the discount rate \$0.039 (Linux, US East, 1-year light utilization).
Since the Google traces only span one month, we proportionally shorten the
on-demand billing cycle from one hour to one minute, and the reservation period
from 1 year to 6 days (i.e., $24 \times 365 = 8760 \mbox{ minutes} = 6
\mbox{ days}$) as well.

\subsection{Evaluations of Online Algorithms}


We start by evaluating the performance of online algorithms without any {\em a
priori} knowledge of user demands.

{\bf Benchmark Online Algorithms:} We compare our online deterministic and
randomized algorithms with three benchmark online strategies. The first is {\em
All-on-demand}, in which a user never reserves and operates all workloads with
on-demand instances. This algorithm, though simple, is the most common strategy
in practice, especially for those users with time-varying workloads
\cite{aws-case-study}. The second algorithm is {\em All-reserved}, in which all
computational demands are served via reservations. The third online algorithm
is the simple extension to the Bahncard algorithm proposed in \cite{bahncard}
(see Sec.~\ref{sec:bahncard}), and is referred to as {\em Separate} because
instances are reserved separately. All three benchmark algorithms, as well as
the two proposed online algorithms, are carried out for each user in the Google
traces. All the incurred costs are {\bf {\em normalized to All-on-demand}}.

{\bf Cost Performance:} We present the simulation results in
Fig.~\ref{fig:cost}, where the CDF of the normalized costs are given,
grouped by users with different demand fluctuation levels.  We see in
Fig.~\ref{fig:all-fluc} that when applied to all 933 users, both the
deterministic and randomized online algorithms realize significant cost savings
compared with all three benchmarks. In particular, when switching from
All-on-demand to the proposed online algorithms, more than 60\% users cut their
costs. About 50\% users save more than 40\%. Only 2\% incur
slightly more costs than before. For users who switch from All-reserved to our
randomized online algorithms, the improvement is even more substantial. As shown in
Fig.~\ref{fig:all-fluc}, cost savings are almost guaranteed, with 30\%
users saving more than 50\%.  We also note that Separate, though generally
outperforms All-on-demand and All-reserved, incurs more costs than our online
algorithms, mainly due to its ignorance of reservation correlations.

We next compare the cost performance of all five algorithms at different demand
fluctuation levels. As expected, when it comes to the extreme cases, 
All-on-demand is the best fit for Group 1 users whose demands are known to be
highly busty and sporadic (Fig.~\ref{fig:high-fluc}), while  All-reserved
incurs the least cost for Group 3 users with stable workloads
(Fig.~\ref{fig:low-fluc}). These two groups of users, should they know their demand
patterns, would have the least incentive to adopt advanced instance reserving
strategies, as naively switching to one option is already optimal. However,
even in these extreme cases, our online algorithms, especially the randomized
one, remain highly competitive, incurring only slightly higher cost.


However, the acquisition of instances is not always a black-and-white choice
between All-on-demand and All-reserved. As we observe from
Fig.~\ref{fig:med-fluc}, for Group 2 users, a more intelligent reservation
strategy is essential, since naive algorithms, either All-on-demand or
All-reserved, are always highly risky and can easily result in skyrocketing
cost. Our online algorithms, on the other hand, become the best choices in this
case, outperforming all three benchmark algorithms by a significant margin.

Table~\ref{tbl:cost} summarizes the average cost performance for each user
group. We see that, in all cases, our online algorithms remain highly
competitive, incurring near-optimal costs for a user.

\begin{table}[tp]
   \centering
   \renewcommand{\arraystretch}{0.95}
   \footnotesize
   \caption{Average cost performance (normalized to All-on-demand).}
   \vspace{-2mm}
   \begin{tabular}{|l||l|l|l|l|}
     \hline
     {\bf Algorithm} & {\bf All users} & {\bf Group 1} & {\bf Group 2} & {\bf Group 3}\\
     \hline
     All-reserved & 16.48 & 48.99 & 1.25 & 0.61 \\
     \hline
     Separate & 0.88 & 1.01 & 1.02 & 0.71 \\
     \hline
     Deterministic & 0.81 & 1.00 & 0.89 & 0.67 \\
     \hline
     Randomized & 0.76 & 1.02 & 0.79 & 0.63 \\
     \hline
   \end{tabular}
   \label{tbl:cost}
   \vspace{-2mm}
\end{table}

\subsection{The Value of Short-Term Predictions}

\begin{figure}[tp]
  \subfloat[Cost CDF]
  {
      \label{fig:deter-pred-cdf}
      \includegraphics[width=0.48\linewidth]{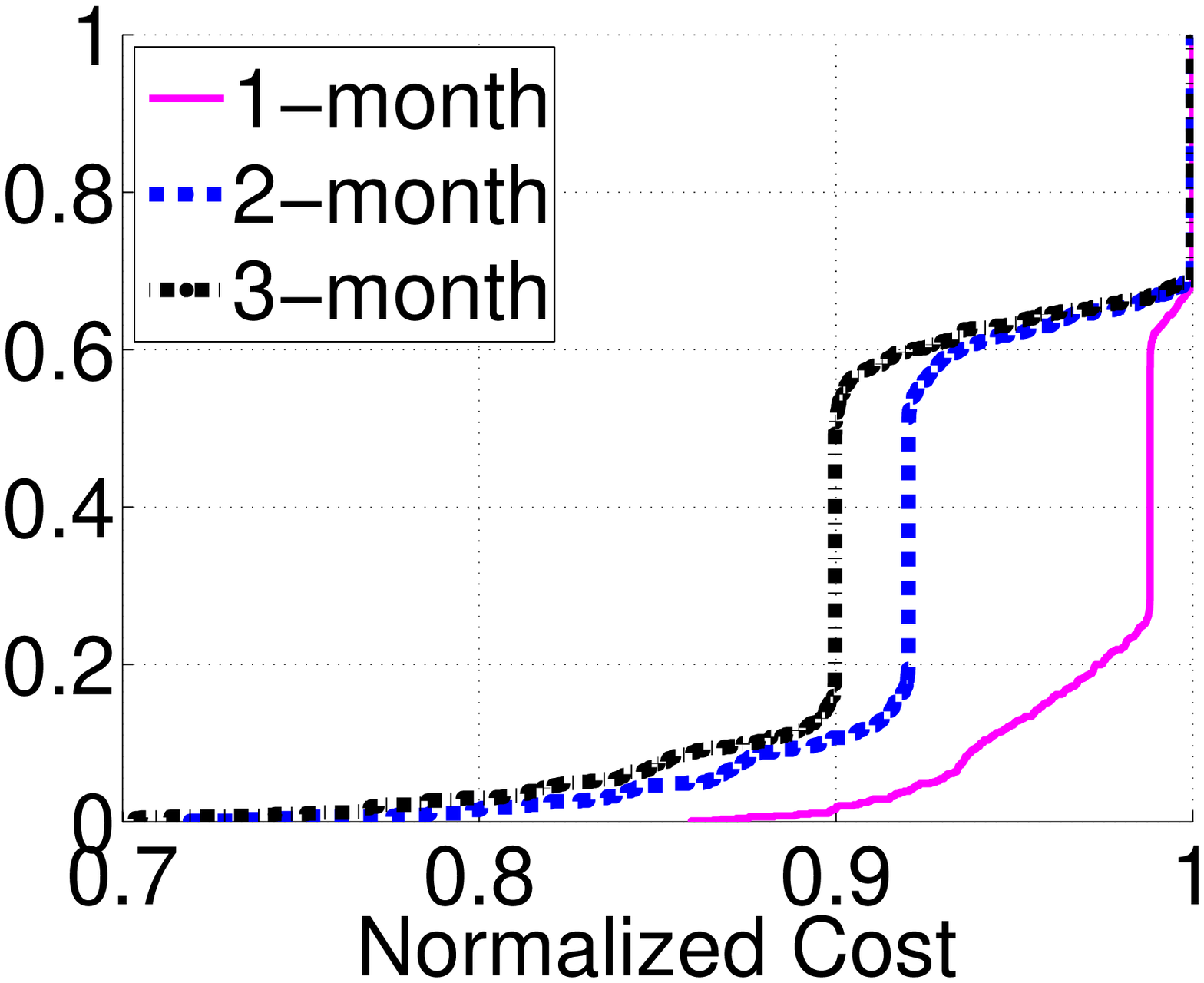}
  }
  \hspace{-3mm}
  \subfloat[Average cost in different user groups]
  {
      \label{fig:deter-pred-bar}
      \includegraphics[width=0.48\linewidth]{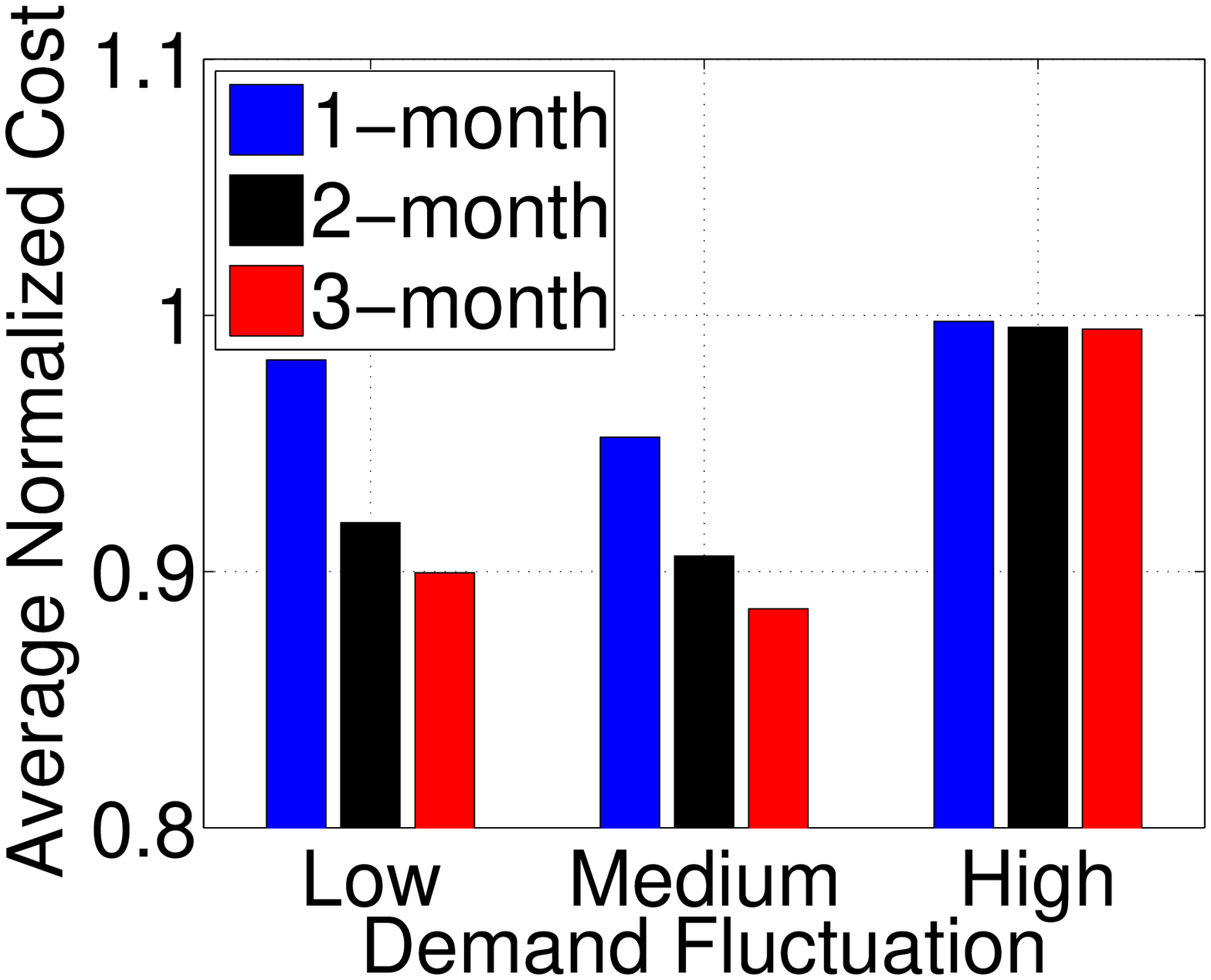}
  }
  \caption{
    Cost performance of the deterministic algorithm with various prediction windows.
    All costs are normalized to the online deterministic algorithm
    (Algorithm~\ref{alg:deter}) without any future information.
  }
  \label{fig:deter-cost-pred}
\end{figure}

\begin{figure}[tp]
  \subfloat[Cost CDF]
  {
      \label{fig:rand-pred-cdf}
      \includegraphics[width=0.48\linewidth]{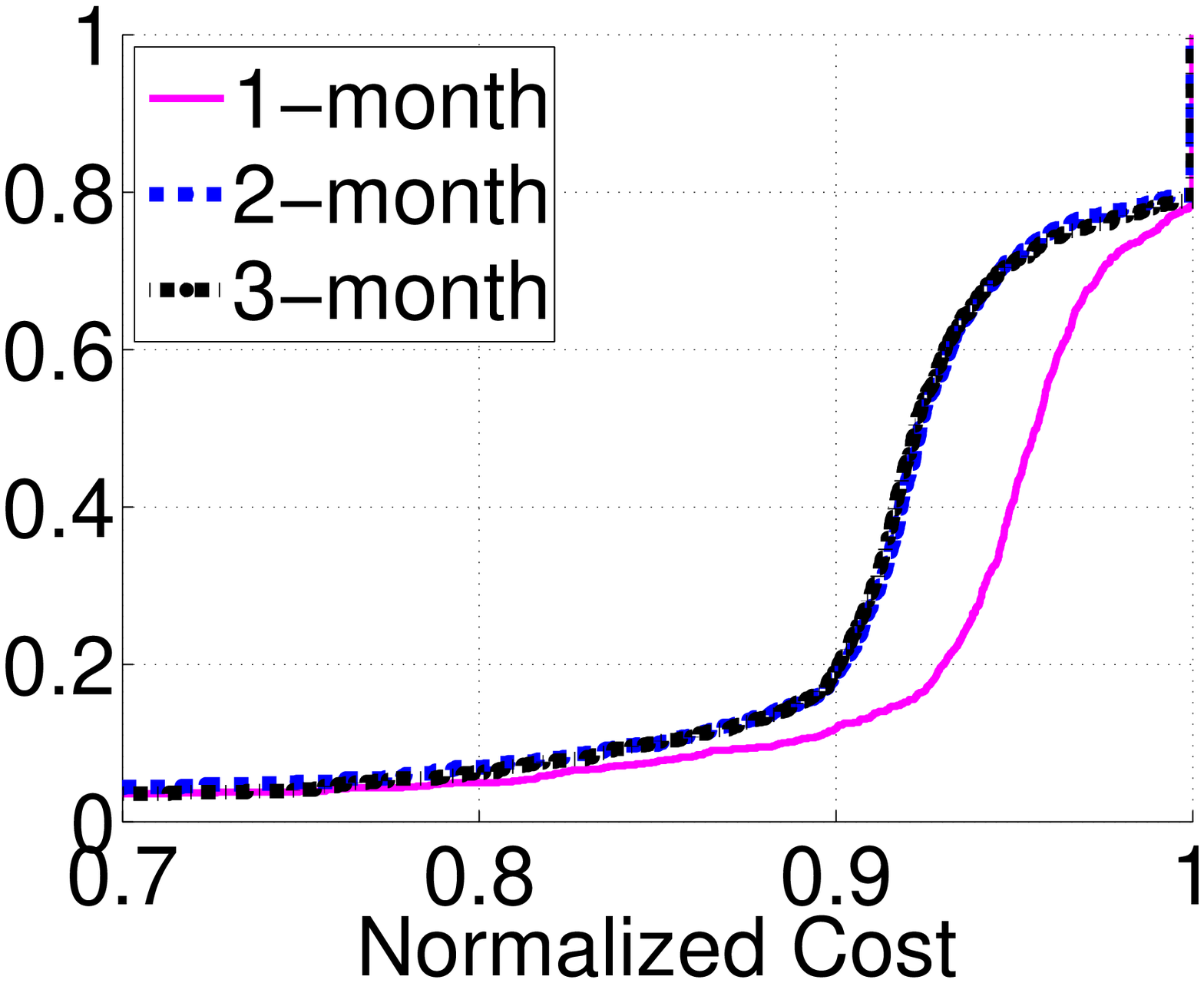}
  }
  \hspace{-3mm}
  \subfloat[Average cost in different user groups]
  {
      \label{fig:rand-pred-bar}
      \includegraphics[width=0.48\linewidth]{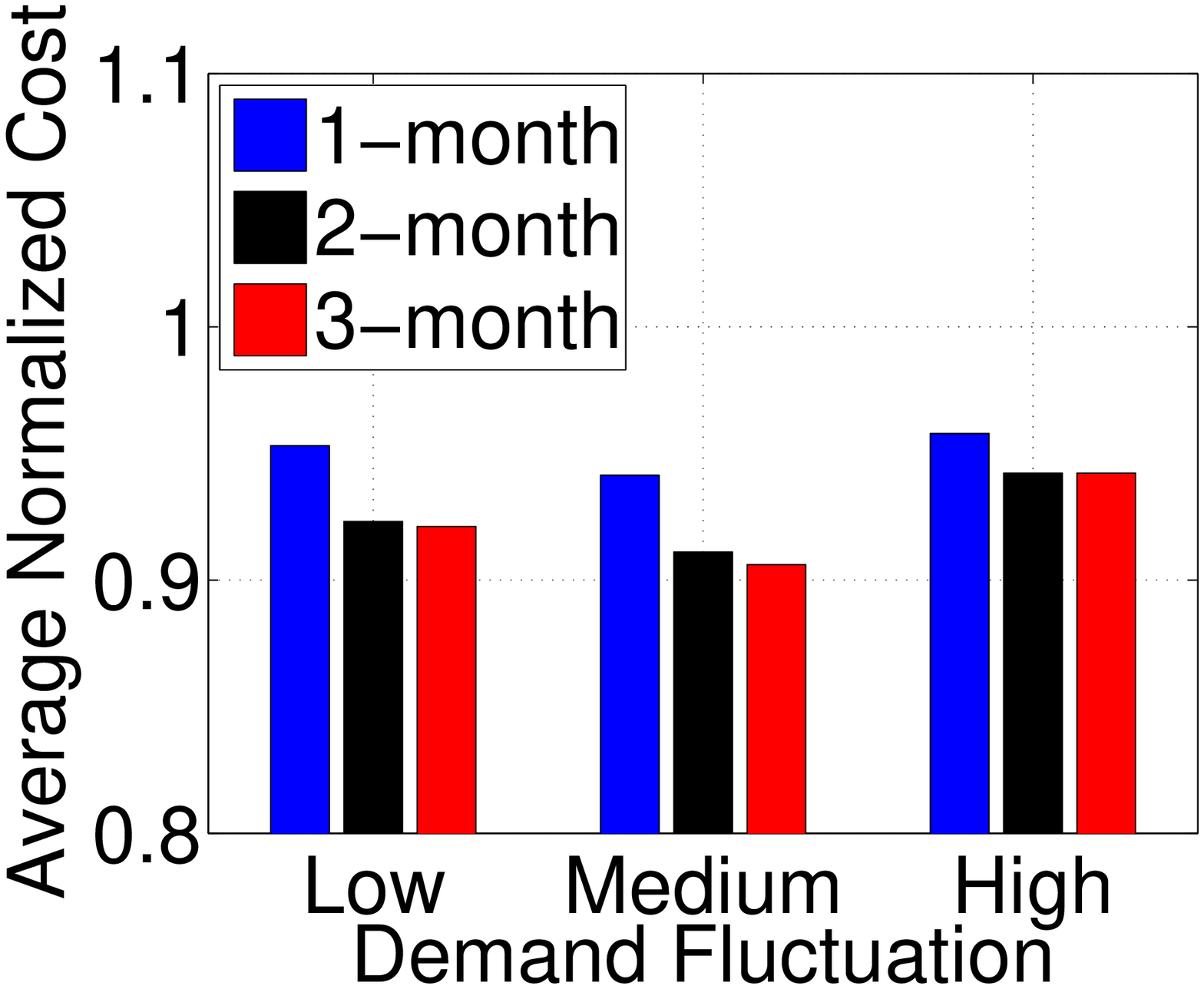}
  }
  \caption{
    Cost performance of the randomized algorithm with various prediction
    windows. All costs are normalized to the online randomized algorithm
    (Algorithm~\ref{alg:deter}) without any future information.
  }
  \label{fig:rand-cost-pred}
  \vspace{-.1in}			
\end{figure}

While our online algorithms perform sufficiently well without knowledge of
future demands, we show in this section that more cost savings are realized by
their extensions when short-term demand predictions are reliable. In
particular, we consider three prediction windows that are 1, 2, and 3 months
into the future, respectively. For each prediction window, we run both the
deterministic and randomized extensions (i.e., Algorithm~\ref{alg:deter-pred}
and \ref{alg:rand-pred}) for each Google user in the traces, and compare their
costs with those incurred by the online counterparts without future knowledge
(i.e., Algorithm~\ref{alg:deter} and \ref{alg:rand}).
Figs.~\ref{fig:deter-cost-pred} and \ref{fig:rand-cost-pred} illustrate the
simulation results, where all costs are {\em normalized to Algorithm~\ref{alg:deter}
and \ref{alg:rand}, respectively}.

As expected, the more information we know about the future demands (i.e.,
longer prediction window), the better the cost performance. Yet, the marginal
benefits of having long-term predictions are diminishing. As shown in
Figs.~\ref{fig:deter-pred-cdf} and \ref{fig:rand-pred-cdf}, long prediction
windows will not see proportional performance gains. This is especially the
case for the randomized algorithm, in which knowing the 2-month future demand
{\em a priori} is no different from knowing 3 months beforehand. 

Also, we can see in Fig.~\ref{fig:deter-pred-bar} that for the deterministic
algorithm, having future information only benefits those users whose demands
are stable or with medium fluctuation. This is because the deterministic online
algorithm is almost optimal for users with highly fluctuating demands (see
Fig.~\ref{fig:high-fluc}), leaving no space for further improvements.  On the
other hand, we see in Fig.~\ref{fig:rand-pred-bar} that the benefits of knowing
future demands are consistent for all users with the randomized algorithm.

\section{Related Work}
\label{sec:related}

On-demand and reserved instances are the two most prominent pricing options
that are widely supported in leading IaaS clouds \cite{ec2pricing,
elastichosts, gogrid}. Many case studies \cite{aws-case-study} show that
effectively combining the use of the two instances leads to a significant cost
reduction.

There exist some works in the literature, including both algorithm design
\cite{hong11, bodenstein11, Wang13a} and prototype implementation
\cite{verm11}, focusing on combining the two instance options in a cost
efficient manner. All these works assume, either explicitly or implicitly, that
workloads are statistically stationary in the long-term future and can be
accurately predicted {\em a priori}. However, it has been observed that in real
production applications, ranging from enterprise applications to large
e-commerce sites, workload is highly variable and statistically non-stationary
\cite{stewart07, singh10}.  Furthermore, most workload prediction schemes,
e.g., \cite{Chen08a, Guenter11a, Goiri12a}, are only suitable for predictions over a very
short term (from half an hour to several hours). Such limitation is also
shared by general predicting techniques, such as ARMA \cite{time-series} and
GARCH models \cite{boller86}.  Some long-term workload prediction schemes
\cite{urgaon05, gmach07}, on the other hand, are reliable only when demand
patterns are easy to recognize with some clear trends. Even in this case, the
prediction window is at most days or weeks into the future \cite{urgaon05},
which is far shorter than the typical span of a reservation period (at least
one year in Amazon EC2 \cite{ec2pricing}).  All these factors significantly
limit the practical use of existing works.


Our online strategies are tied to the online algorithm literature
\cite{borodin98}. Specifically, our instance reservation problem captures a
class of rent-or-buy problems, including the ski rental problem
\cite{ski-rental}, the Bahncard problem \cite{bahncard}, and the TCP
acknowledgment problem \cite{tcp-ack}, as special cases when a user demands no
more than one instance at a time. In these problems, a customer obtains {\em a
	single item} either by paying a repeating cost (renting) per usage or
by paying a one-time cost (buying) to eliminate the repeating cost. A customer
makes {\em one-dimensional} decisions only on the timing of buying. Our problem
is more complicated as a user demands {\em multiple instances} at a time and
makes {\em two-dimensional} decisions on both the timing and quantity of its
reservation.  A similar ``multi-item rent-or-buy'' problem has also been investigated in
\cite{Lu12b}, where a dynamic server provisioning problem is considered
and an online algorithm is designed to dynamically turn
on/off servers to serve time-varying workloads with a minimum energy cost.  It
is shown in \cite{Lu12b} that, by dispatching jobs to servers that are idle or
off the most recently, the problem reduces to a set of independent ski rental
problems.  Our problem does not have such a separability structure and cannot
be equivalently decomposed into independent single-instance reservation
(Bahncard) problems, mainly due to the possibility of time multiplexing
multiple jobs on the same reserved instance. It is for this reason that the
problem is challenging to solve even in the offline setting.

Besides instance reservation, online algorithms have also been applied to
reduce the cost of running a file system in the cloud. The recent work
\cite{Khanafer13a} introduces a {\em constrained ski-rental problem} with extra
information of query arrivals (the first or second moment of the distribution),
proposing new online algorithms to achieve
improved competitive ratios. \cite{Khanafer13a} is orthogonal
to our work as it takes advantage of additional demand information to make
rent-or-buy decisions for a single item.

\section{Concluding Remarks and Future Work}
\label{sec:conclusion}

Acquiring instances at the cost-optimal commitment level for time-varying
workloads is critical for cost management to lower IaaS service costs. In
particular, when should a user reserve instances, and how many instances should
it reserve? Unlike existing reservation strategies that require knowledge of
the long-term future demands, we propose two online algorithms, one
deterministic and another randomized, that dynamically reserve instances
without knowledge of the future demands. We show that our online algorithms incur
near-optimal costs with the best possible competitive ratios, i.e., $2-\alpha$
for the deterministic algorithm and $e/(e-1+\alpha)$ for the randomized
algorithm. Both online algorithms can also be easily extended to cases when
short-term predictions are reliable. Large-scale simulations driven by 40 GB
Google cluster-usage traces further indicate that significant cost savings are
derived from our online algorithms and their extensions, under the prevalent
Amazon EC2 pricing.

One of the issues that we have not discussed in this paper is the combination of
different types of reserved instances with different reservation periods and
utilization levels. For example, Amazon EC2 offers 1-year and 3-year reserved
instances with light, medium, and high utilizations. Effectively combining
these reserved instances with on-demand instances could further reduce instance
acquisition costs. We note that when a user demands no more than one instance at
a time and the reservation period is infinite, the problem reduces to {\em
	Multislope Ski Rental} \cite{Lotker08a}. However, it remains unclear if
and how the results obtained for Multislope Ski Rental could be extended to
instance acquisition with multiple reservation options. 


\bibliographystyle{IEEEtran}
\bibliography{main}

\appendices

\section{Proof of Lemma~\ref{lem:resv-num}}
\label{app:deter}

\begin{figure}[tb]
  \centering
  \includegraphics[width=0.40\textwidth]{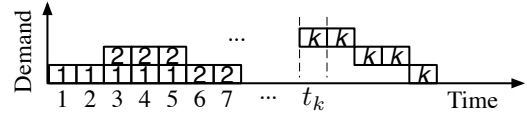}
  \vspace{-.1in}
  \caption{A reservation can be pictorially represented as a reservation strip. For example,
  reservation 2 is represented as $(2,2,2,1,1)$ with $t_2 = 3$.
  }
  \vspace{-.1in}
  \label{fig:resv-strip}
\end{figure}

We first present a general description for a reserved instance, based on which
we reveal the connections between reservations in $A_\beta$ and OPT.

Suppose an algorithm reserves $n$ instances 1, 2, \dots, $n$ at time $t_1 \le
t_2 \le \dots \le t_n$, respectively. Also suppose at time $t$, the active
reservations are $i, i+1, \dots, j$. Let demand $d_t$ be divided into levels,
with level 1 being the bottom. Without loss of generality, we can serve demand
at level 1 with reservation $i$, and level 2 with $i+1$, and so on. This gives
us a way to describe the use of a reserved instance to serve demands.
Specifically, the reservation $k$ is active from $t_k$ to $t_k+\tau-1$ and
is described as a $\tau$-tuple $(l_{t_k}^k, l_{t_k + 1}^k, \dots,
l_{t_k + \tau - 1}^k)$, where $l_t^k$ is the demand level that $k$ will serve at
time $t = t_k, \dots, t_k + \tau - 1$. Such a tuple can be pictorially
represented as a {\em reservation strip} and is depicted in
Fig.~\ref{fig:resv-strip}.

\begin{figure}[tb]
  \centering
  \includegraphics[width=0.40\textwidth]{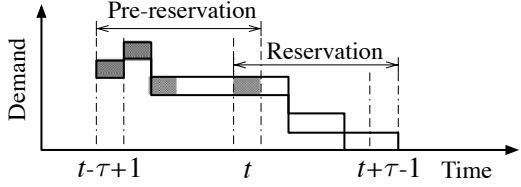}
  \vspace{-.1in}
  \caption{Illustration of a decision strip for a reserved instance in Fig.~\ref{fig:deter}.
  }
  \vspace{-.1in}
  \label{fig:dec-strip}
\end{figure}

We next define a {\em decision strip} for every reserved instance in $A_\beta$
and will show its connections to the reservation strips in OPT. Suppose an
instance is reserved at time $t$ in $A_\beta$, leading $x_i$'s to update in
line~\ref{lne:future} and \ref{lne:phantom} of Algorithm~\ref{alg:deter}. We
refer to the bottom figure of Fig.~\ref{fig:deter} and define the decision
strip for this reservation as the region between the original $x$ curve (the
solid line) and the newly updated one (the dotted line).
Fig.~\ref{fig:dec-strip} plots the result, where the shaded area is derived
from the top figure of Fig.~\ref{fig:deter}. Such a decision strip captures
critical information of a reserved instance in $A_\beta$. As shown in
Fig.~\ref{fig:dec-strip}, the first $\tau$ times of the strip, referred to as
the {\em pre-reservation part}, reveal the reason that causes this reservation:
serving the shaded area on demand incurs more costs than the break-even point
$\beta$.  The last $\tau$ times are exactly the reservation strip of this
reserved instance in $A_\beta$, telling how it will be used to serve demands.
We therefore denote a decision strip of a reservation $k$ as a
$(2\tau-1)$-tuple $\mathbf{l}_k = (l_{t_k - \tau + 1}^k, \dots, l_{t_k}^k,
\dots, l_{t_k + \tau - 1}^k)$, where $l_t^k$ is the demand level of strip $k$ at
time $t = t_k - \tau + 1, \dots, t_k + \tau - 1$. The pre-reservation part is
$(l_{t_k - \tau + 1}^k, \dots, l_{t_k}^k)$, while the reservation strip is
$(l_{t_k}^k, \dots, l_{t_k + \tau - 1}^k)$.

We now show the relationship between decision strips of $A_\beta$ and
reservation strips of OPT. Given the demand sequence, Algorithm $A_\beta$ makes
$n_\beta$ reservations, each corresponding to a decision strip. OPT reserves
$n_\mathrm{OPT}$ instances, each represented as a reservation strip.  We say a
decision strip of $A_\beta$ {\em intersects} a reservation strip of OPT if the
two strips share a common area when depicted. The following lemma establishes
their connections.

\begin{lemma}
  \label{lem:single-resv-strip}
  A decision strip of $A_\beta$ intersects at least a reservation strip of OPT.
\end{lemma}

{\bf Proof:}
  We prove by contradiction. Suppose there exists a decision strip of $A_\beta$
  that intersects no reservation strip of OPT. In this case, the demand in the
  pre-reservation part (the shaded area in Fig.~\ref{fig:dec-strip}) are served
  on demand in OPT, incurring a cost more than the break-even point (by the
  definition of $A_\beta$). This implies that the cost of OPT can be further
  lowered by serving these pre-reservation demands (the shaded area) with a
  reserved instance, contradicting the definition of OPT.
\qed

\begin{corollary}
  \label{cor:mult-resv-strip}
  Any two decision strips of $A_\beta$ intersect at least two reservation
  strips of OPT, one for each.
\end{corollary}

{\bf Proof:}
  We prove by contradiction. Suppose there exist two decision strips of
  $A_\beta$, one corresponding to reservation $i$ made at time $t_i$ and another to reservation
  $j$ made at $t_j$, $i<j$, that intersect {\em only one} reservation strip of OPT. (By
  Lemma~\ref{lem:single-resv-strip}, any two decision strips must intersect at
  least one reservation strip of OPT.) By the analyses of
  Lemma~\ref{lem:single-resv-strip}, the reservation strip of OPT intersects
  the pre-reservation parts of both decision strip $i$ and $j$. It suffices to
  consider the following two cases.

\begin{figure}[tb]
  \centering
  \subfloat[Strip $i$ and $j$ do not overlap in time.]
  {
      \label{fig:no-overlap}
      \includegraphics[width=0.45\textwidth]{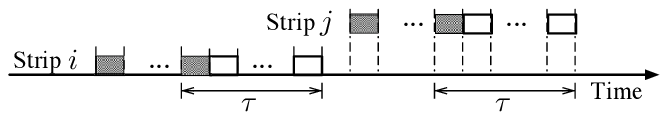}
  }
  \vspace{-.1in}\\
  \subfloat[Strip $i$ and $j$ overlap in time.]
  {
      \label{fig:overlap}
      \includegraphics[width=0.45\textwidth]{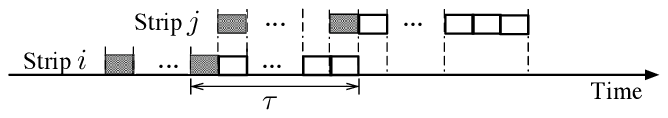}
  }
  \caption{Two cases of strip $i$ and $j$. They either overlap in time or not.
  }
  \vspace{-.1in}
  \label{fig:strip-case}
\end{figure}

  {\em Case~1:} Strip $i$ and $j$ do not overlap in time, i.e., $t_i + \tau - 1 <
  t_j - \tau + 1$. As shown in Fig.~\ref{fig:no-overlap}, having a reservation strip of
  OPT intersecting the pre-reservation parts of both $i$ and $j$ requires a
  reservation period that is longer than $\tau$, which is impossible.

  {\em Case~2:} Strip $i$ and $j$ overlap in time. In this case, the
  reservation strip of OPT only intersects strip $i$. To see this, we refer to
  Fig.~\ref{fig:overlap}. Because $i < j$, strip $i$ is depicted below $j$,
  i.e., $l_t^i < l_t^j$ for all $t$ in the overlap. Suppose the reservation
  strip of OPT intersects decision strip $j$ at time $t$. Clearly, $t$ must be
  in the overlap and $l_t^i < l_t^j$. This implies that at time $t$ in OPT, OPT serves
  the demand at a lower level $l_t^i$ by an on-demand instance while serving a
  higher level $l_t^j$ via a reservation, which contradicts the definition of
  the reservation strip.
 \qed

With Lemma~\ref{lem:single-resv-strip} and Corollary~\ref{cor:mult-resv-strip},
we see that the $n_\beta$ decision strips of $A_\beta$ intersect at least
$n_\beta$ reservation strips of OPT, indicating that $n_\beta \le
n_\mathrm{OPT}$.

\section{Proof Sketch of Lemma~\ref{lem:rand}}
\label{app:rand}

{\bf Statement 1:} Following the notations used in the proof of
Proposition~\ref{prop:deter-comp}, let $S$ be the total costs of demands when
priced at the on-demand rate, and $\mathrm{Od}(A_z)$ be the on-demand costs incurred by
algorithm $A_z$. It is easy to see
\begin{equation}
  \label{eq:az-opt}
  \left\{
    \arraycolsep=1.5pt
    \begin{array}{lcl}
      C_{A_z} & = & n_z + (1 - \alpha) \mathrm{Od}(A_z) + \alpha S ; \\
      C_\mathrm{OPT} & = & n_\mathrm{OPT} + (1 - \alpha) \mathrm{Od}(\mathrm{OPT}) + \alpha S .
    \end{array}
  \right.
\end{equation}
Plugging (\ref{eq:az-opt}) into (\ref{eq:cost-az}), we see that it is equivalent to proving
\begin{equation}
  \label{eq:od-az}
  \mathrm{Od}(A_z) \le \mathrm{Od}(\mathrm{OPT}) + z n_\mathrm{OPT} - E_z ~.
\end{equation}

Denote by $\mathrm{Od}(A_z \backslash \mathrm{OPT}) := \sum_{t=1}^T (o_{z,t} -
o_t^*)^+ p$ the on-demand costs incurred by $A_z$ that are not incurred by OPT.
With similar arguments as we made for (\ref{eq:od-cost}), we see
$\mathrm{Od}(A_z \backslash \mathrm{OPT}) \le z n_\mathrm{OPT}$. We therefore derive
\begin{align}
  \mathrm{Od}(A_z) & = \mathrm{Od}(\mathrm{OPT}) + \mathrm{Od}(A_z \backslash \mathrm{OPT})
  - E_z \nonumber \\
  & \le \mathrm{Od}(\mathrm{OPT}) + z  n_\mathrm{OPT}
  - E_z ~, \nonumber
\end{align}
which is exactly (\ref{eq:od-az}).
\qed

{\bf Statement 2:} For any given demands $\{d_t\}$, let $L(n,z)$ be the minimum, over all
purchase decisions $D = \{ r_t, o_t \}$ with $n$ reserved instances, of the
on-demand cost that has been incurred in $D$ but has not been incurred in
$A_z$, i.e.,
\begin{equation}
  \begin{split}
    L(n,z) & := \mbox{$ \inf_{ \{ r_t, o_t \} } \sum_{t=1}^T (o_t - o_{z,t})^+ p$} \\
    \mbox{s.t.} & \quad \mbox{$\sum_{t=1}^T r_t = n$,} \\
    & \quad \mbox{$o_t + \sum_{i=t-\tau+1}^t r_i \geq d_t$,} \\
    & \quad \mbox{$o_t, r_t \in \{0,1,2,\dots\}$, $t=1,\dots,T$.}
  \end{split}
\end{equation}

We show that for any $u > v \ge z$,
\begin{equation}
  \label{eq:L}
  L(n_u, z) \ge (v - z)(n_v - n_u) + L(n_v, z) ~.
\end{equation}
To see this, let $D_u = \{r_{u,t}, o_{u,t}\}$ be the purchase decision that
leads to $L(n_u,z)$ and define decision strips for $A_v$ and reservation strips
for $D_u$ similarly as we did in Appendix~\ref{app:deter}.  With similar
arguments of Corollary~\ref{cor:mult-resv-strip}, we see that a reservation
strip of $D_u$ intersects at most one decision strip of $A_v$. As a result,
among all $n_v$ decision strips of $A_v$, there are at least $n_v - n_u$ ones
that do not intersect any reservation strips of $D_u$.  We arbitrarily choose
$n_v - n_u$ such decision strips and denote their collections as $B$. Each of these decision strips contains a pre-reservation part
with an on-demand cost $v$\footnote{This is true when $p \ll 1$.}, of which at most $z$ is also incurred on demand in
$A_z$ (by definition of $A_z$).  As a result, an on-demand cost of at least $(v
- z)(n_v - n_u)$ is incurred in both $D_u$ and $A_v$ that is not incurred in
$A_z$, i.e., 
\begin{equation}
  \sum_{t=1}^T (\min\{o_{v,t}, o_{u,t} \} - o_{z,t})^+ p \ge (v-z)(n_v - n_u)~,
\end{equation}
where $o_{v,t}$ is the number of on-demand instances launched by $A_v$ at time $t$.
We now reserve a new instance at the starting time of each decision strip in $B$.
Adding to the existing reservations made by $D_u$, we have $D' = D_u \cup B$ as
new reserving decisions with $n_v$ reserved instances (because $|B| = n_v - n_u$). Therefore
\begin{align}
  L(n_u, z) & = \mathrm{Od}(D_u \backslash A_z) \nonumber \\
  & \ge \mathrm{Od}(D' \backslash A_z) + (v - z)(n_v - n_u) \nonumber \\
  & \ge L(n_v, z) + (v - z)(n_v - n_u) ~,
\end{align}
where the second inequality holds due to the definition of $L(n_v, z)$ and the
fact that $D'$ consists of $n_v$ reserved instances.

The rest of the proof follows the framework of \cite{tcp-ack}.
Taking $u = v + \ud v$ and integrating from $z$ to $w$, for any $z < w \le \beta$, we have
\begin{equation}
  L(n_w, z) - L(n_z, z) \ge \int_z^w n_v \ud v - (w - z) n_w ~.
\end{equation}
Observing that $L(n_z, z) = 0$, and that $n_v \le n_w$ for $v > w$, we have
\begin{equation}
  L(n_w, z) \ge \int_z^\beta n_v \ud v - (\beta - z) n_w ~.
\end{equation}
Taking $n_w = n_\mathrm{OPT}$ and noting that $E_z \ge L(n_\mathrm{OPT}, z)$ yields the statement.
\qed

{\bf Statement 3:}
  By (\ref{eq:az-opt}) and noting that $\mathrm{Od}(\mathrm{OPT}) \ge E_z$, we
  have 
  \begin{align}
    C_\mathrm{OPT} & \ge n_\mathrm{OPT} + \alpha S + (1 - \alpha) E_z \nonumber \\
    & \ge n_\mathrm{OPT} + E_z~,
  \end{align}
  where the second inequality is derived by noting $S
  \ge E_z$. Letting $z \to 0$ and plugging (\ref{eq:ez}) establishes the
  statement.
\qed

\end{document}